\documentclass[iop,useAMS,usenatbib,preprint,letterpaper]{emulateapj}
\usepackage{amsmath}
\usepackage{graphicx}
\usepackage[normalem]{ulem}
\shorttitle{Effect of Magnetic Fields on the CMF}
\shortauthors{Bailey \& Basu}

\begin{document}

\title{The Effect of Magnetic Fields and Ambipolar Diffusion on Core Mass Functions}

\author{Nicole D. Bailey and Shantanu Basu}
\affil{Department of Physics and Astronomy, University of Western Ontario\\
 1151 Richmond Street, London, Ontario, N6A 3K7}
\email{nwityk@uwo.ca (NDB); basu@uwo.ca (SB)}

\begin{abstract}
Linear analysis of the formation of protostellar cores in planar magnetic interstellar clouds yields information about length scales involved in star formation. Combining these length scales with various distributions of other environmental variables, (i.e., column density and mass-to-flux ratio) and applying Monte Carlo methods allow us to produce synthetic core mass functions (CMFs) for different environmental conditions. Our analysis shows that the shape of the CMF is directly dependent on the physical conditions of the cloud. Specifically, magnetic fields act to broaden the mass function and develop a high-mass tail while ambipolar diffusion will truncate this high-mass tail. In addition, we analyze the effect of small number statistics on the shape and high-mass slope of the synthetic CMFs. We find that observed core mass functions are severely statistically limited, which has a profound effect on the derived slope for the high-mass tail. 
\end{abstract}


\section{Introduction}

Observations of the stellar initial mass function (IMF) and the core mass function (CMF) show similarities in the shape and high mass slope of these two functions \citep[][among others]{Motte1998, TS1998, Johnstone2000, Alves2007, NWT2007, Simpson2008, Enoch2008, Sadavoy2010}. As such, much theoretical effort has been invested in order to explain these similarities. Various different approaches to this problem have been explored, including analytic and numerical studies which invoke gravitational fragmentation or accretion \citep{Silk1995, Inutsuka2001, BJ2004}, turbulence \citep{Padoan1997, PN2002, BP2006, Hennebelle2008, Hennebelle2009}, independent stochastic processes \citep{Larson1973, Elmegreen1997} and magnetic fields \citep{Dib2008}, among others. Results of these studies vary from those which seem to agree with the fiducial Salpeter form, $dN/d\log M \propto M^{-\alpha}$ where $\alpha = 1.35$ is the value of the Salpeter slope, to those that do not. 

The high mass slope of the IMF was initially derived by  \citet{Salpeter1955} and later improved upon by \citet{Kroupa2002} and \citet{Chabrier2003a, Chabrier2003b, Chabrier2005}. Despite variations in observed and theoretically derived IMF slope values, it is often assumed that the shape and high mass slope of the IMF and CMF are identical and universal. From a theoretical view, such a one-to-one correspondence between these two functions implies that high-mass cores beget high-mass stars and likewise for low-mass cores. The need for extensive simulations of how a complex of cores turns into a cluster of stars is simplified tremendously if it is assumed that each core will collapse into a single star with some mass loss to account for the mass shift between the CMF and IMF. 

The underlying tenet of universality is that all star-forming regions are the same and undergo the same process to form stars, however recent observations and simulations have started to reveal cracks in this assumption. In a study of the effect of turbulence on the formation of the CMF, \citet{Hennebelle2008, Hennebelle2009} find that comparisons between their IMF and observations for different cloud conditions suggest that star formation should predominantly occur in clouds five times denser than characterized by \citet{Larson1981}. This led them to question the universality of the IMF since, as they say, choosing different cloud parameters would lead to a different CMF/IMF. Several recent studies of the IMF also tend to disagree with the assumed universality. Observations of different star clusters in both the Milky Way and the Large Magellanic Cloud (LMC) show a wide scatter of slopes: $\alpha = 0.5 - 2.0$ \citep{Elmegreen1999}. A survey of high mass slope values for different stars (i.e., cluster stars versus association stars versus field stars) yields a wide range of values; $\alpha = 2.0 - 4.0$ for extreme field stars to $\alpha = 1.5-2.0$ for cluster stars \citep{Elmegreen1997}. Further to this, \citet{Elmegreen1999} shows that through stochastic fractal sampling of a cloud, the derived IMF slopes can vary from $\alpha$ as low as 1.0 to as high as 1.7. \citet{Clark2007} note that if the lifetime of a more massive core is longer than a less massive one, the slope of the CMF should be shallower in order to obtain the IMF. Finally, \citet{Zaritsky2012} show that there may be evidence for two distinct stellar IMFs that depend on the age and metallicity of the cluster in question. Based on the above evidence and arguments, it is not clear why one should insist on using $\alpha = 1.35$ as the universal slope for both the CMF and IMF.

The majority of the work in this area has focused on the effects of turbulence within the molecular clouds on the formation and shape of the CMF. Research which considers the effect of magnetic fields and ambipolar diffusion on the CMF is sparse. \citet{KM2009} used the results of a non-ideal MHD linear analysis of a partially ionized sheet \citep{mor91, CB2006} to generate a broad CMF, assuming ambipolar-diffusion initiated core formation. Their model assumed subcritical to critical initial conditions with a uniform distribution of mass-to-flux ratios between 0.1 and 1.0 times the critical value for gravitational instability (see Section 5 for more discussion of their model).

In this study, we use the results of the linear analysis of a partially ionized sheet along with a lognormal distribution of initial column density and various distributions of mass-to-flux ratio. We explore both subcritical and supercritical initial conditions. Mildly supercritical initial conditions are the most likely to lead to massive core formation, as seen in e.g., Figure 2 of \citet{CB2006}. Furthermore, we use a lognormal distribution of column densities, as expected in molecular clouds on both theoretical grounds for a turbulent medium \citep{Padoan1997} and from observations \citep{Kain2009}. The aim of this paper is two fold. In the first part we show the effects of a magnetic field on the shape of the CMF. Starting from an assumption of lognormal column density probability we show the broadening effect of neutral-ion drift via ambipolar diffusion and differing mass-to-flux ratio distributions. In the second part, we address the inherent limitations of observed core mass functions, i.e. sample size and bin size. Specifically, we aim to compare small sample synthetic CMFs to large sample synthetic CMFs to show effect of small number statistics on the observed features of the CMF. In Section 2 we outline our model and methods for constructing our synthetic CMFs. Section 3 shows the results for the different distribution models considered. Section 4 shows the effect of small number statistics and the variance in derived analytic slopes. Finally Sections 5 and 6 give our discussion and conclusions. 

\section{Synthetic Core Mass Functions}

To better understand the effects of the environment on the shape and peak of the core mass function, we produce synthetic CMFs (synCMFs) based upon varying physics and properties of molecular clouds. These include the column density ($\sigma_{n,0}$), ionization fraction ($\chi_{i}= \log[n_{e}/n_{H}]$), mass-to-flux ratio ($\mu_{0}$), and neutral ion-collision time ($\tau_{ni}$). The synCMFs are produced by randomly sampling predefined column density and mass-to-flux ratio distributions (where applicable) and using a preferred fragmentation length scale to calculate the core mass. We choose to use such methods due to the random nature of molecular cloud properties. This allows us to statistically determine the shape of the CMF for a wide range of randomly chosen $\sigma_{n} - \mu_{0}$ pairs.

\subsection{Physical Model}
\label{physmodel}
We consider the formation of cores and the resulting CMF within ionized, isothermal, interstellar molecular clouds. These clouds are modelled as planar sheets with infinite extent in the $x$- and $y$- directions and a local vertical half thickness $Z$. The nonaxisymmetric equations and formulations of our assumed model have been described in detail in several papers \citep{CB2006, Basu2009a, Basu2009b, BB2012}. For this work we consider three models: nonmagnetic, flux-frozen magnetic field and a magnetic field with ambipolar diffusion.

The key ingredient to this analysis is the assumed length scale for the core. This length scale for collapse can be derived through linear analysis. The nonaxisymmetric equations of \citet{CB2006} and \citet{Basu2009a, Basu2009b} include the effect of ambipolar diffusion. This is quantified by the timescale for collisions between ions bound to the magnetic field and free neutral particles. This timescale is 
\begin{equation} 
\tau_{ni} = 1.4 \left(\frac{m_i +m_{H_2}}{m_i} \right) \frac{1}{n_i\langle\sigma w\rangle_{iH_2}}.
\end{equation} 
Here, $m_{i}$ is the ion mass, $n_{i}$ is the number density of ions and $\langle\sigma w\rangle_{iH_2}$ is the neutral-ion collision rate. The typical atomic and molecular species within a molecular cloud are singly ionized Na, Mg and HCO which have a mass of 25 amu. Assuming collisions between H$_{2}$ and HCO$^+$, the neutral-ion collision rate is $1.69\times 10^{-9}$ cm$^{3}$ s$^{-1}$ \citep{MM1973}. Collisions between neutrals and ions transfer information about the magnetic field to the neutral particles. The threshold for whether a region of a molecular cloud is stable or unstable to collapse is given by the mass-to-flux ratio of the background reference state 
\begin{equation}
\mu_{0} \equiv 2\pi G^{1/2}\frac{\sigma_{n,0}}{B_{\rm ref}},
\end{equation}
where $(2\pi G^{1/2})^{-1}$ is the critical mass-to-flux ratio for gravitational collapse in the adopted model and $B_{\rm ref}$ is the magnetic field strength of the reference state. Regions with $\mu_{0} < 1$ are defined as subcritical, regions with $\mu_{0} > 1$ are defined to be supercritical and regions with $\mu_{0} \sim 1$ are transcritical.

A dispersion relation for the governing magnetohydrodynamic equations can be found via linear analysis \citep{CB2006, Basu2009b, BB2012} . Here we follow the analysis as described in \citet{BB2012}. For a model with ambipolar diffusion, the resulting dispersion relation is 

\begin{eqnarray}
\nonumber(\omega&+&i\theta)(\omega^{2} - C_{\rm eff,0}^2 k^2 + 2\pi G \sigma_{n,0}k) \\
                &=& \omega(2\pi G\sigma_{n,0}k\mu^{-2}_{0} + k^2V_{A,0}^2)
\label{fulldisp}
\end{eqnarray}
where
\begin{equation}
\theta = \tau_{ni,0}(2\pi G\sigma_{n,0}k\mu^{-2}_{0} + k^2V_{A,0}^2).
\end{equation}
Here, $\omega$ is the angular frequency of the perturbations, $\tau_{ni,0}$ is the initial neutral-ion collision time, $k$ is the wavenumber in the $z$-direction, $V_{A,0}$ is the Alfv\'en speed, where 
\begin{equation}
V_{A,0}^{2} \equiv \frac{B_{\rm ref}^{2}}{4\pi\rho_{n,0}} = 2\pi G\sigma_{n,0}\mu_{0
}^{-2}Z_{0},
\end{equation}
$Z_{0}$ is the initial half-thickness of the sheet, and $C_{\rm eff,0}$ is the local effective sound speed, such that
\begin{equation}
C_{\rm eff,0}^{2} = \frac{\pi}{2}G\sigma_{n,0}^{2}\frac{[3P_{\rm ext}+(\pi/2)G\sigma_{n,0}^{2}]}{[P_{\rm ext}+(\pi/2)G\sigma_{n,0}^{2}]^{2}}c_{s}^{2}.
\end{equation}
Here,  $c_{s} = (k_{B}T/m_{n})^{1/2}$ is the isothermal sound speed, $k_{B}$ is the Boltzmann constant, $T$ is the temperature in Kelvins and $m_n$ is the mean mass of a neutral particle ($m_{n} = 2.33$ amu). For this analysis, we assume a temperature $T = 10$ K and a normalized external pressure $\tilde{P}_{\rm ext} \equiv 2P_{\rm ext}/\pi G \sigma_{n,0}^{2} = 0.1$.

\begin{center}
\begin{figure}
\centering
\includegraphics[width=0.5\textwidth,angle=0]{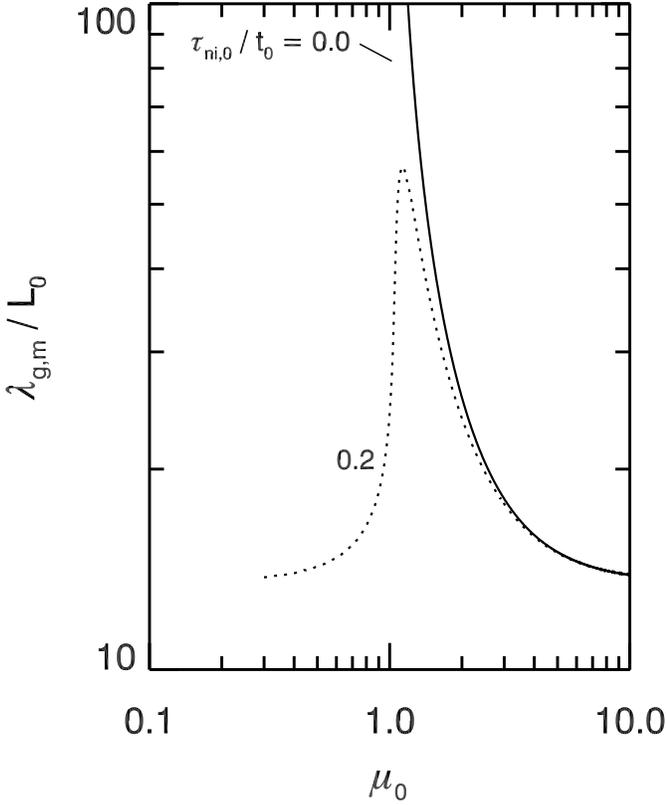}
\caption{Wavelength with minimum growth time as a function of initial mass-to-flux ratio. Displayed curves are for $\tau_{ni,0}/t_{0} = 0$ (solid curve, flux freezing) and $\tau_{ni,0}/t_{0} = 0.2$ (dotted curve). }
\label{lammufig}
\end{figure}
\end{center}
In the limit of flux freezing, $\tau_{ni,0} \rightarrow 0$, which gives the reduced dispersion relation
\begin{equation}
\omega^{2} + 2\pi G \sigma_{n,0}k(1-\mu^{-2}_{0})- k^{2}(C_{\rm eff,0}^2+V_{A,0}^2) = 0.
\label{disprel}
\end{equation}
The gravitationally unstable mode corresponds to one of the roots of $
\omega^2 < 0$ and occurs for $\mu_{0} > 1$. The growth time for this mode can be written as
\begin{equation}
\tau_{g} = \frac{\lambda}{2\pi[G\sigma_{n,0}(1-\mu_0^{-2})(\lambda - \lambda_{MS})]^{1/2}}
\label{eqn:taug}
\end{equation}
for $\lambda \geq \lambda_{MS}$, where
\begin{equation}
\lambda_{MS} =\frac{{C}_{\rm eff,0}^2 + V_{A,0}^2}{G\sigma_{n,0}(1 - \mu_{0}^{-2})}.
\label{lammu}
\end{equation}
The length scale corresponding to the minimum growth time is $\lambda_{g,m} = 2\lambda_{MS}$. This is the length scale used to produce our synCMFs for models with flux freezing. The variation of this length scale as a function of $\mu_{0}$ is shown by the solid line in Figure~\ref{lammufig}. For the case with no magnetic field, Equation~\ref{lammu} reduces down to the thin disk equivalent of the Jeans length,
\begin{equation}
\lambda_{J} = \frac{C_{\rm eff}^{2}}{G\sigma_{n,0}}.
\end{equation}
Again, the length scale corresponding to the minimum growth time is $\lambda_{g,m,J} = 2\lambda_{J}$, which is the scale used in our nonmagnetic model. 

\begin{center}
\begin{figure}
\centering
\includegraphics[width=0.5\textwidth,angle=0]{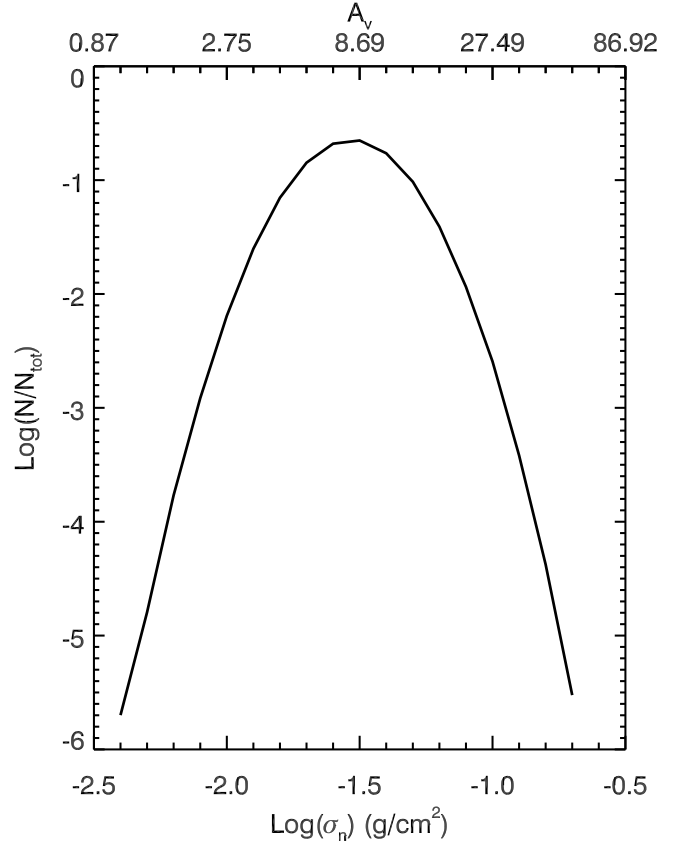}
\caption{Model lognormal column density distribution.}
\label{densdist}
\end{figure}
\end{center}

The addition of ambipolar diffusion complicates the process somewhat. In these cases, the gravitationally unstable mode corresponds to one of the roots of the full dispersion relation (Equation~\ref{fulldisp}). However since it is a cubic function, there is no simple expression to describe these roots. Therefore, each length scale is computed numerically. The value of this length scale is related to the degree of ambipolar diffusion i.e., the degree of ionization within the cloud, and the mass-to-flux ratio of the region. Previous studies show that the ionization fraction within a molecular cloud resembles a step function \citep{Ruffle1998,BB2012} such that the outer layers are highly ionized due to UV photoionization while ionization of denser inner regions is primarily due to cosmic rays. For this study, we choose to fix the neutral-ion collision time to the dimensionless value $\tau_{ni,0}/t_{0} = 2\pi G\sigma_{n,0}\tau_{ni,0}/c_{s} =  0.2$ ; a value typical of the denser inner regions where most cores are likely to form \citep[][and references within]{Basu2009a}. This corresponds to an ionization fraction $\chi_{i} = 5.2\times 10^{-8}$ at a neutral column density $\sigma_{n,0} = 0.023 \rm{~g~cm^{-2}}$. Figure~\ref{lammufig} (dotted line) shows the relation between the collapse length scale and the mass-to-flux-ratio for this neutral-ion collision time. By fixing the neutral-ion collision time, our ambipolar diffusion models have only two free parameters, the column density and mass-to-flux ratio distributions. Our choices for these two parameters are discussed in the following sections. 

\begin{center}
\begin{figure}
\centering
\includegraphics[width=0.23\textwidth,angle=0]{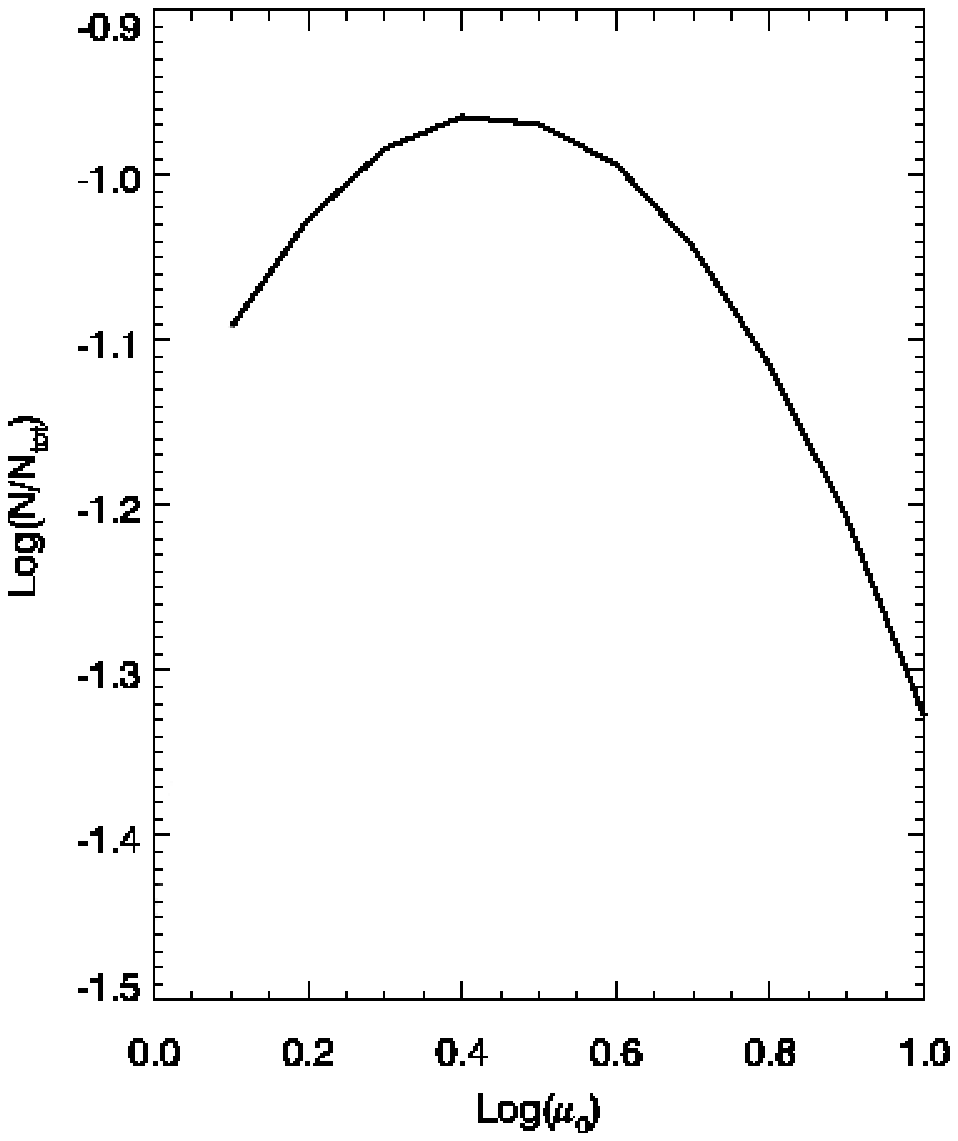}
\includegraphics[width=0.22\textwidth,angle=0]{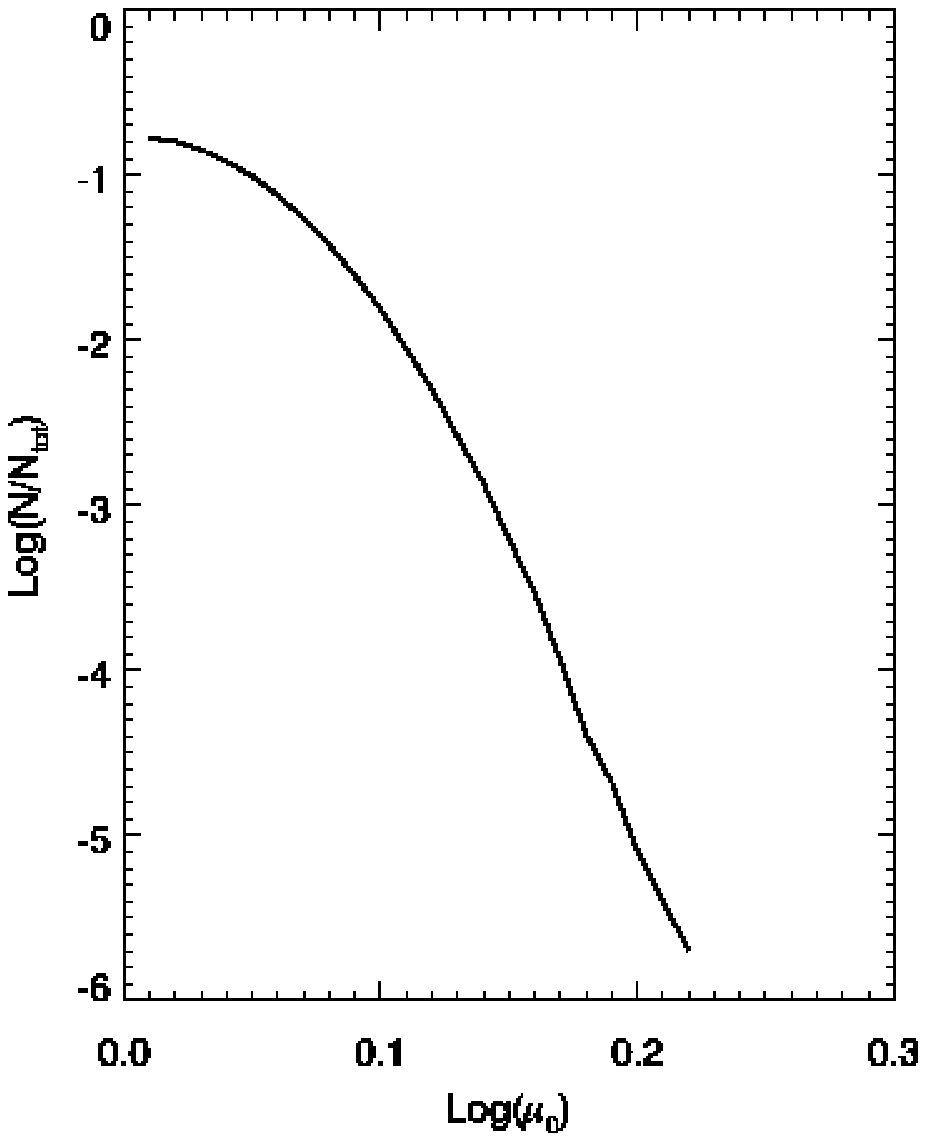}
\caption{Model mass-to-flux distributions for flux freezing models. Left: Broad Lognormal Distribution (FF2). Right: Narrow Lognormal distribution (FF3).}
\label{mudistff}
\end{figure}
\end{center}
\subsection{Column Density Distribution}
\label{densdistsec}
A survey of column density \textbf{$\sigma_{n}$} distributions within various molecular clouds shows that they generally exhibit log-normal distributions either with or without a high density tail \citep{Kain2009}. Correlation of these different shapes with the conditions within the clouds suggest that regions with a pure lognormal distribution tend to be quiescent while those with high density tails show signs of active star formation.  

Since the aim of this paper is to investigate the shape of the core mass function as an initial condition for star formation, we choose a simple lognormal distribution as shown in Figure~\ref{densdist}. This plot shows the distribution as a function of both the column density ($\sigma_{n}$, lower axis) and the visual extinction ($A_{v}$, upper axis). Following the prescription of \citet{Pineda2010}, the conversion from visual extinction to column density is achieved by combining the ratio of $H_{2}$ column density to color excess \citep{Bohlin1978} with the total selective extinction \citep{Whittet2003} to yield a conversion factor $N(H_{2}) = 9.35\times 10^{20} A_{v}~\rm cm^{-2}~mag^{-1}$. Although this conversion is specifically for $H_{2}$, the abundance ratio of CO to $H_{2}$ is $\sim 10^{-4}$ and other molecular contributions are even smaller, so they do not add significantly to the number density of $H_{2}$. Therefore we assume this number density is representative of all species. Assuming a mean molecular weight of 2.33 amu, this translates into a mass column density conversion of the form
\begin{equation} 
\sigma_{n} =3.638\times 10^{-3} A_{v}~\rm g~cm^{-2}~mag^{-1}.
\label{av2sigma}
\end{equation}

The variance and mean ($\sigma^{2}$ and $\mu$) of this distribution were chosen based upon observational information. Previous studies of molecular clouds show visual extinction thresholds for core and star formation to be on the order of $A_{v} = 5$ mag \citep{Johnstone2004, Kirk2006} and $A_{v} = 8$ mag \citep[see][among others]{Johnstone2004,Froebrich2010} respectively. As such, we adopted a mean visual extinction value of 8 magnitudes for our lognormal density distribution. The variance reflects the typical width of the lognormal fits to cloud density functions presented by \citet{Kain2009}.  

\subsection{Mass-to-Flux Ratio Distributions}

Although density/visual extinction maps are fairly commonplace, measurements of magnetic field strengths 
\begin{center}
\begin{figure}
\centering
\includegraphics[width=0.23\textwidth,angle=0]{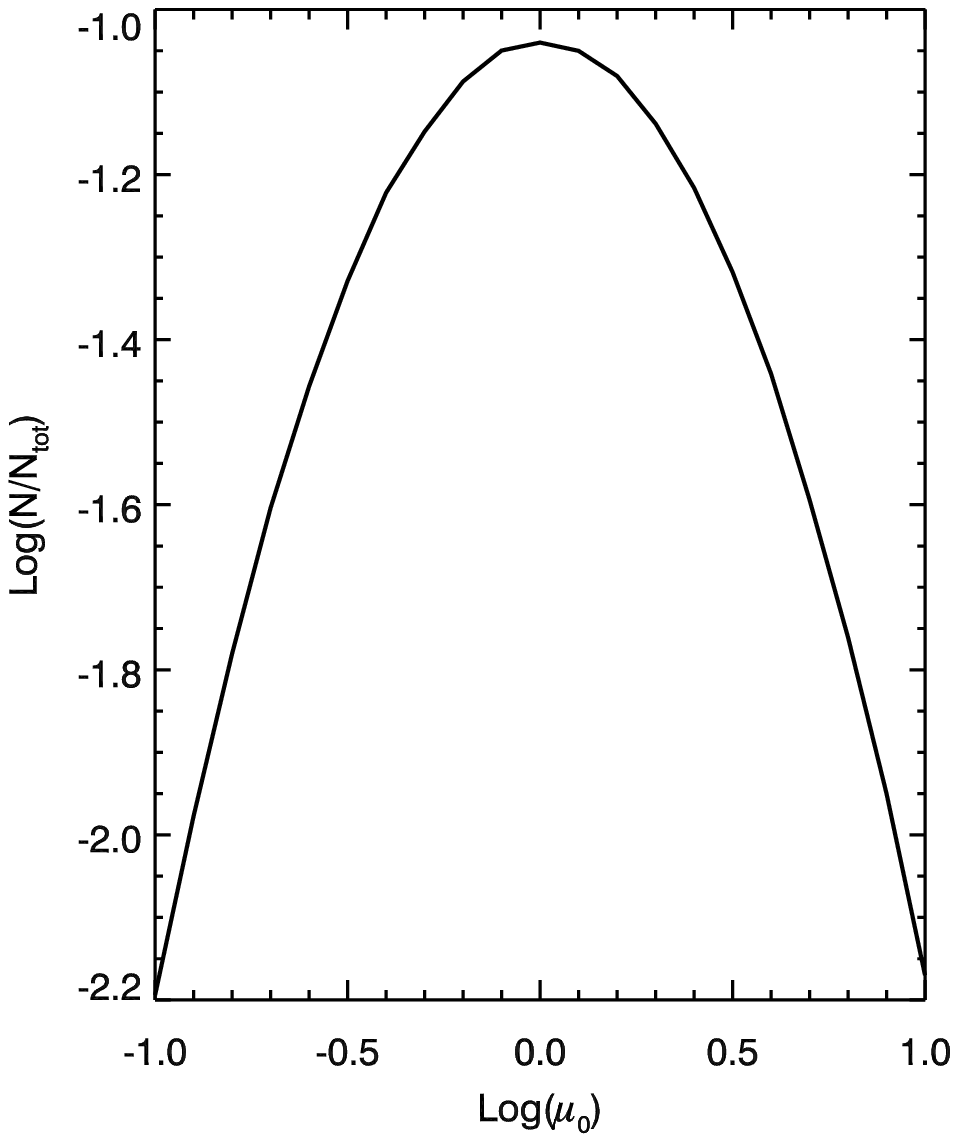}
\includegraphics[width=0.23\textwidth,angle=0]{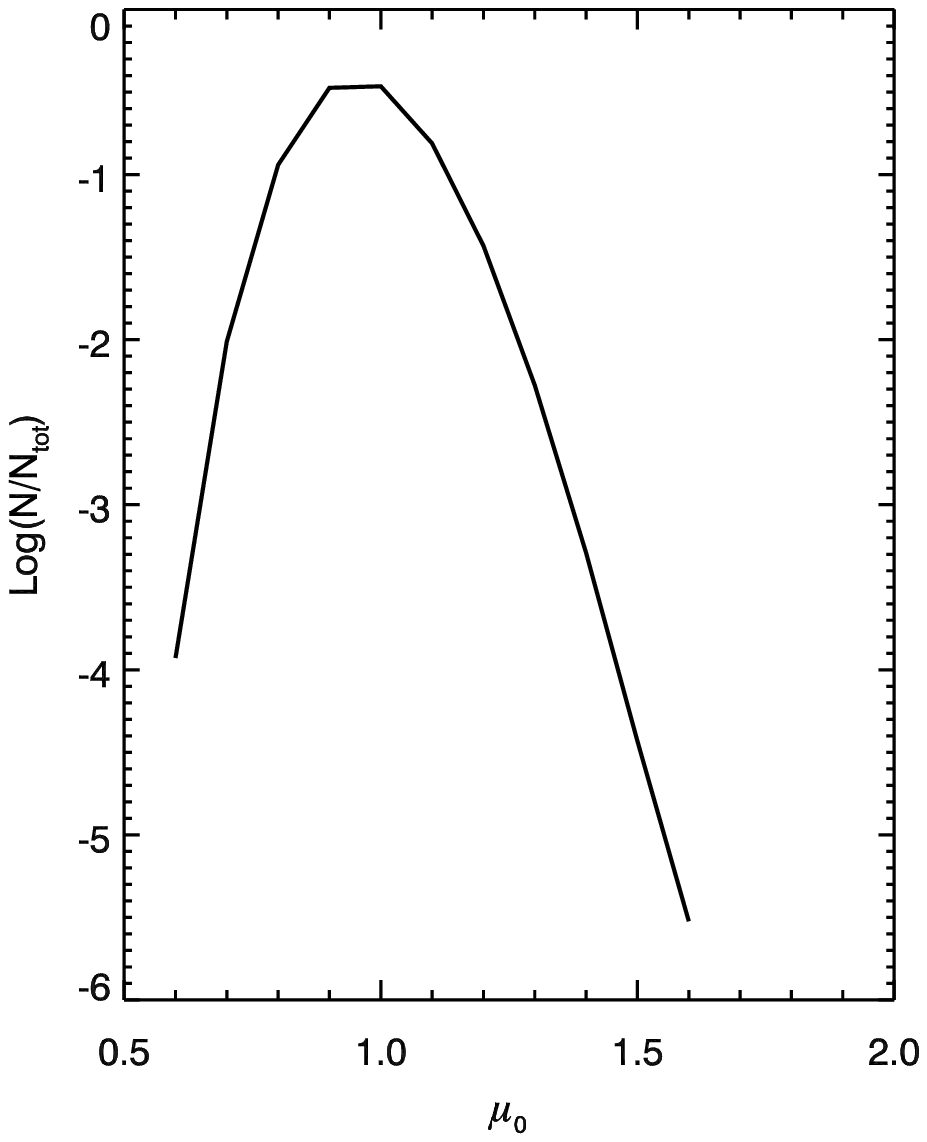}
\caption{Model mass-to-flux distributions for ambipolar diffusion models. Left: Broad Lognormal Distribution (AD4), Right: Narrow Lognormal (AD5).}
\label{mudistAD}
\end{figure}
\end{center}
within molecular clouds are difficult to obtain. Due to limitations in techniques and resolution, studies of magnetic fields within clouds are generally on a more global scale \citep[see][among others]{Crutcher1999, HT2004, TC2008, Falgarone2008, Crutcher2010, Chapman2011} which does not give much insight into the exact nature of $\mu_{0}$ within denser small scale regions. Therefore, the mass-to-flux ratio of specific regions are not generally known, let alone a distribution over an entire cloud. Recent simulations of cloud formation with magnetic fields \citep{VS2011} show that the mass-to-flux ratio distribution seems to exhibit a lognormal shape. On the other hand, analysis of the likelihood of different magnetic field distributions \citep{Crutcher2010} show that the magnetic field strengths for various regions (HI diffuse clouds, OH dark clouds, etc) exhibit a uniform distribution ranging from very small values up to a maximum value. This seems to disagree with the simulations of \citet{VS2011}. With these results in mind, we choose to explore both options (i.e., uniform and lognormal distributions). 

As shown by the linear analysis results presented in \citet{BB2012} and Figure~\ref{lammufig}, the length scale for collapse is dependent on the value of the mass-to-flux ratio. The value of $\mu_{0}$ is selected from a predefined distribution that is independent of the distribution of $\sigma_{n}$. This implies that the magnetic field strength is not constant and varies according to the choices of $\sigma_{n}$ and $\mu_{0}$. The independent sampling of values of $\sigma_{n}$ and $\mu_{0}$ does not then allow for any systematic dependence of one quantity on the other. We believe this is an acceptable first approximation since the initial conditions of the mass-to-flux ratio distribution in a molecular cloud are poorly constrained. We test several possible $\mu_{0}$ distributions in an attempt to determine if the shape of an observed CMF could reveal information about the underlying mass-to-flux ratio distribution. We consider both uniform and lognormal distributions. Figures~\ref{mudistff}~\&~\ref{mudistAD} show the adopted lognormal mass-to-flux ratio distributions for the flux freezing and ambipolar diffusion models respectively. Specifically, all distributions sample the transcritical peak in fragmentation scale, $\lambda_{g,m}$ (see Figure~\ref{lammufig}). The properties of all $\mu_{0}$ distributions considered are given in Table~\ref{allmodels}.

\begin{table*}
\centering
\caption{Model Parameters}
\begin{tabular}{llccc}
\hline
Model Name & $\mu_{0}$ Distribution & Mean ($\mu$) & Variance ($\sigma^{2}$) & $\mu_{0}$ Range\\
\hline
\hline
\multicolumn{5}{c}{Flux Frozen Models}\\
\hline
FF1 & Uniform & - & - &  1.0 - 3.0\\
FF2 & Broad lognormal & 0.01 & 1.0 & 1.0 - 10\\
FF3 & Narrow lognormal &0.01 & 0.01 & 1.0 - 1.5\\
\hline
\multicolumn{5}{c}{Ambipolar Diffusion Models}\\
\hline
AD1 & Subcritical Uniform & - & - & 0.1 - 1.0 \\
AD2 & Supercritical Uniform & - & - & 1.0 - 3.0 \\
AD3 & Uniform & - & - &  0.7 - 3.0\\
AD4 & Broad lognormal & 0.01 & 1.0 & 0.3 - 10\\
AD5 & Narrow lognormal &0.01 & 0.01 & 0.6 - 1.5\\
\hline
\label{allmodels}
\end{tabular}
\end{table*}

\begin{center}
\begin{figure}
\centering
\includegraphics[width=0.20\textwidth,angle=0]{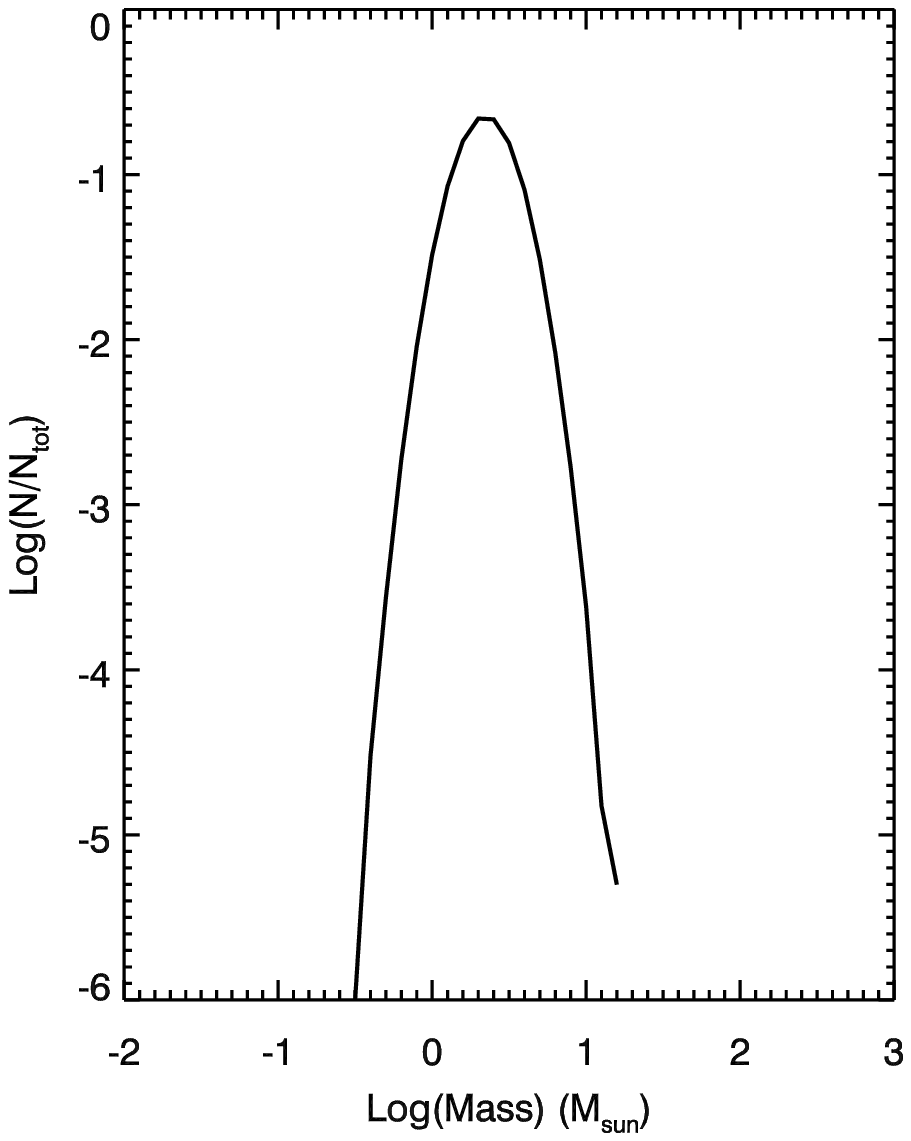}
\includegraphics[width=0.22\textwidth,angle=0]{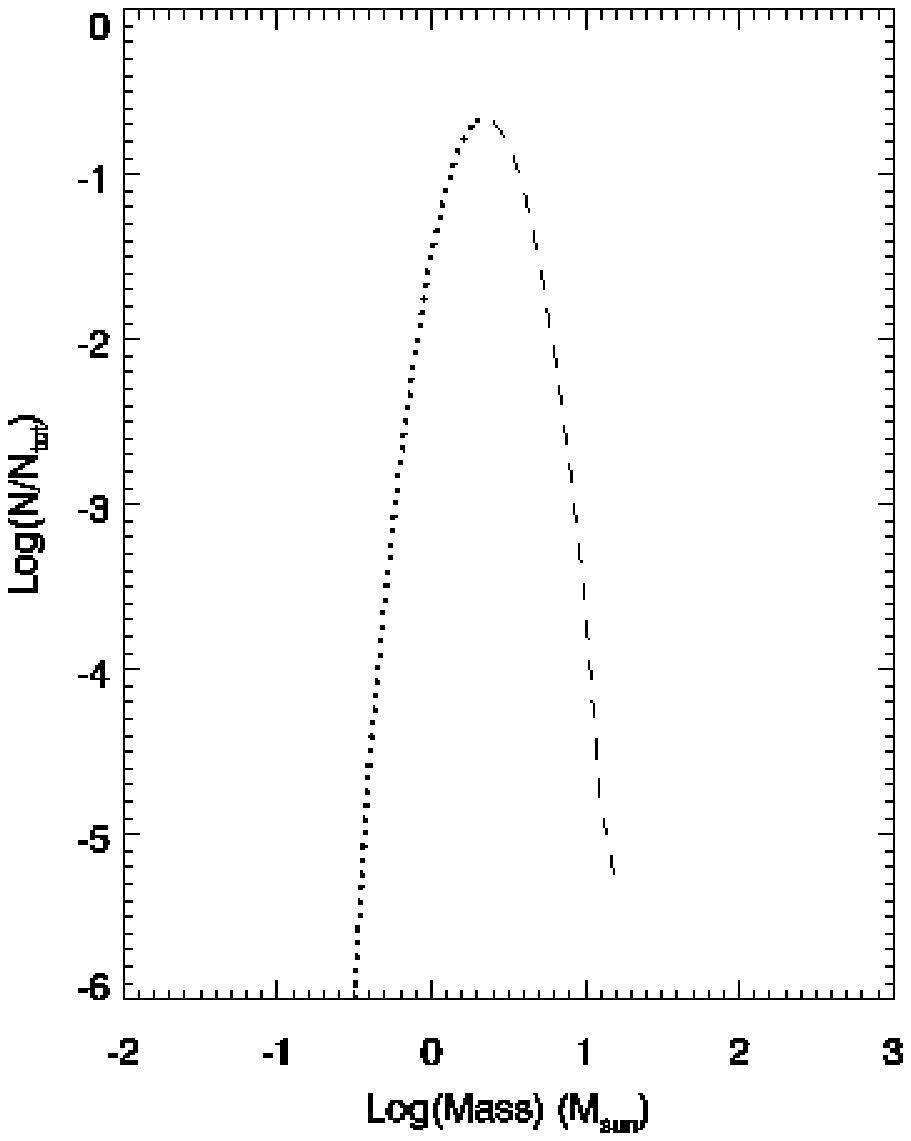}
\caption{Synthetic core mass function for a non magnetic cloud. Left: Total core mass function. Right: Contributions to the core mass function from cores with $A_{v} < 8$ mag (dashed line) and cores with $A_{v} > 8$ mag (dotted line).}
\label{cmfNM}
\end{figure}
\end{center}
\subsection{Producing Synthetic Core Mass Functions}
To produce a synthetic CMF, we randomly sample the column density distribution for the nonmagnetic case and both the column density and mass-to-flux ratio distributions for the magnetic cases. These values are then used to find the preferred length scale for collapse from the linear analysis. Finally, the mass is determined by multiplying the column density by the square of the corresponding length scale. By randomly sampling each model distribution $10^{6}$ times, a synthetic CMF is produced. 

\section{Models and Results}

Our analysis covers several different mass-to-flux ratio distributions and assumptions about the neutral-ion collision time and column density distribution. As stated earlier, the column density distribution is the same for all models (see Figure~\ref{densdist}) and the neutral-ion collision time for the ambipolar diffusion models is set to a normalized value, $\tau_{ni,0}/t_{0}=0.2$. In addition to the models listed in Table~\ref{allmodels}, we also present a nonmagnetic (NM) fiducial case. The following subsections present the results for each model individually. An in depth comparison between all the models and implications regarding observed CMFs will be discussed in Sections~\ref{assess}~\&~\ref{obs} respectively.

\subsection{Non-Magnetic Model}

The nonmagnetic model serves as a baseline for our investigation. The left panel of Figure~\ref{cmfNM} shows the resulting core mass function from this technique. As discussed in Section~\ref{densdistsec}, we choose the peak of our density distribu-
\begin{center}
\begin{figure}
\centering
\includegraphics[width=0.22\textwidth,angle=0]{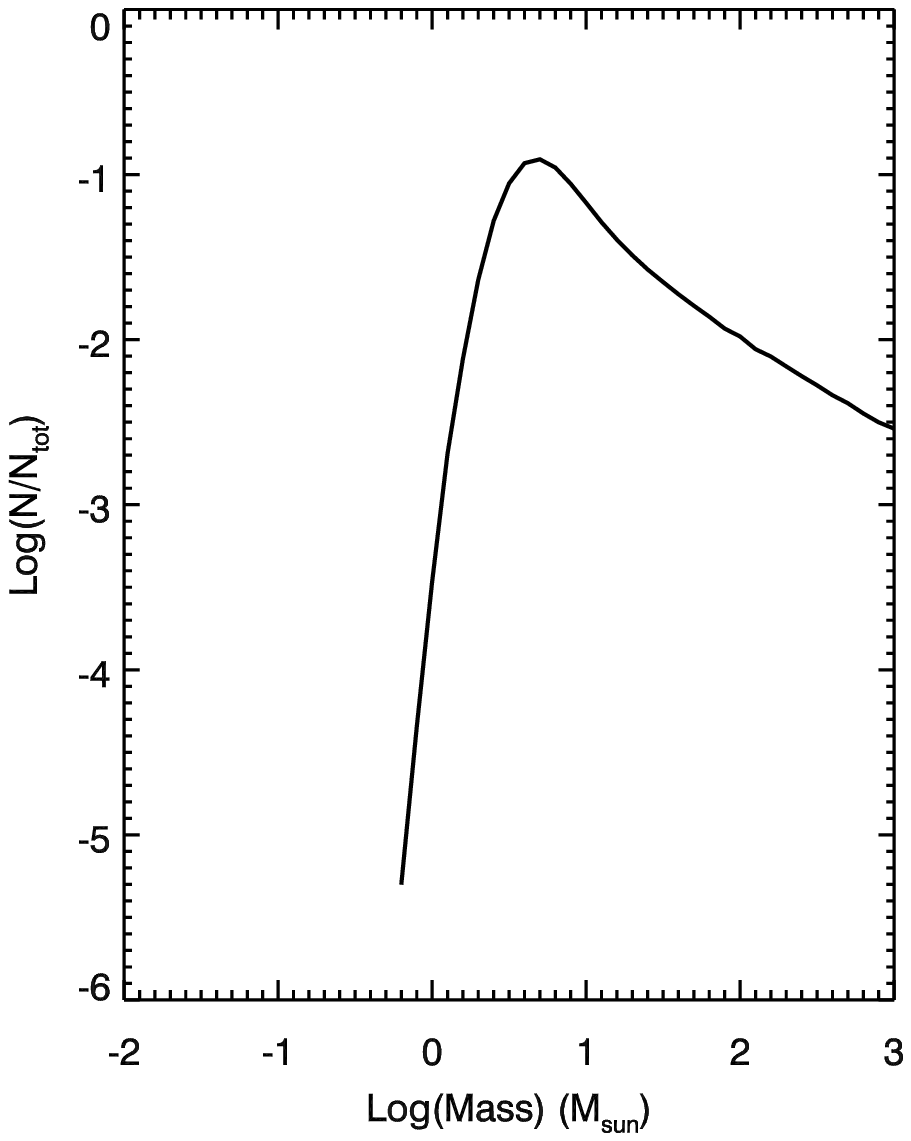}
\includegraphics[width=0.22\textwidth,angle=0]{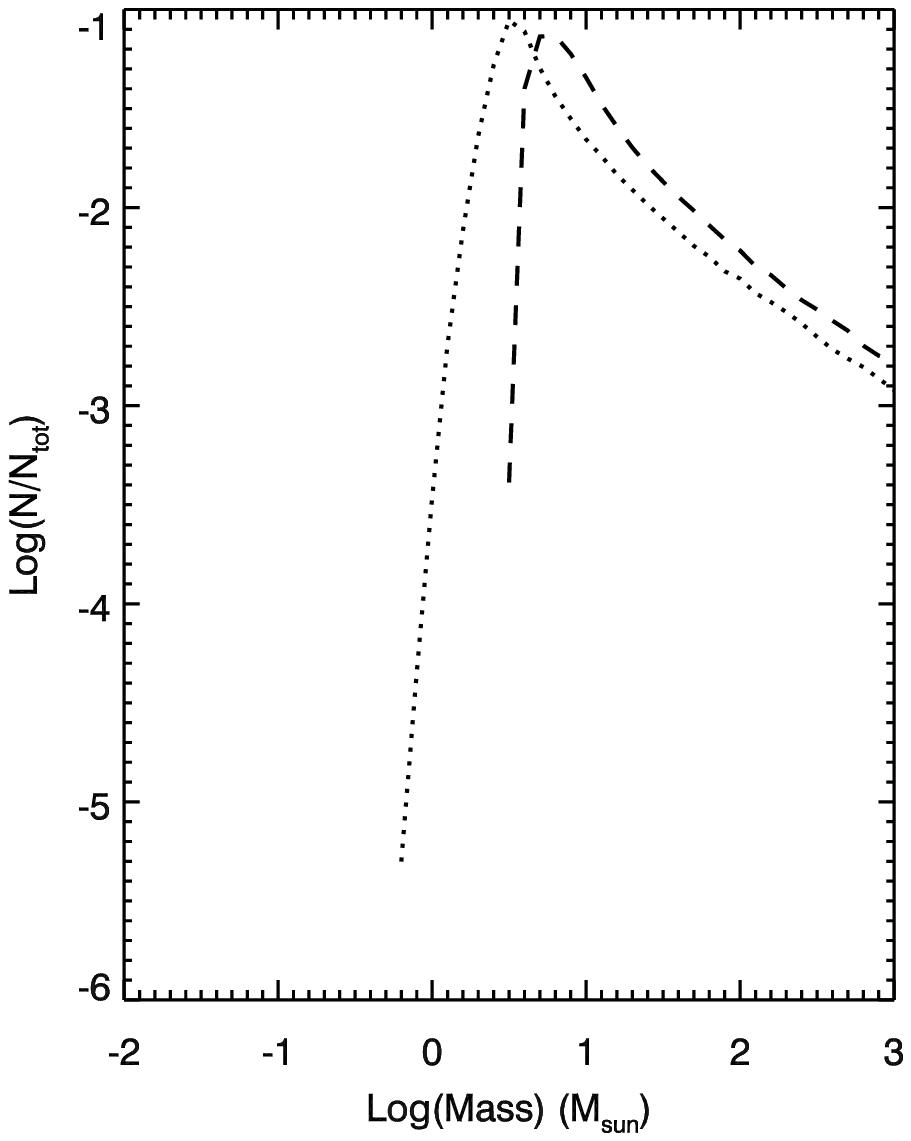}
\caption{Synthetic core mass functions for a flux frozen magnetic cloud assuming a uniformly distributed mass-to-flux ratio (FF1). Left: Total core mass function. Right: Contributions to the core mass function from cores with $A_{v} < 8$ mag (dashed line) and cores with $A_{v} > 8$ mag (dotted line).}
\label{cmfFF1}
\end{figure}
\end{center}
tion to correspond to the apparent visual extinction threshold for the creation of star forming cores; $A_{v} \sim 8$ magnitudes. The right hand panel of Figure~\ref{cmfNM} shows the contributions from high density gas ($A_{v} >8$ mag, dotted line) and low density gas ($A_{v} < 8$ mag, dashed line). As expected from the Jeans theory, the core mass distribution mimics the column density distribution, with high mass cores formed from low density gas and low mass cores formed from high density gas. The distribution of masses for this model peaks at a value of $\log(M/M_{\odot}) = 0.4$ or $M \simeq 2.5 \rm M_{\odot}$ which is consistent with observations \citep{NWT2007}. 

\subsection{Flux Frozen Magnetic Model}
\begin{center}
\begin{figure}
\centering
\includegraphics[width=0.22\textwidth,angle=0]{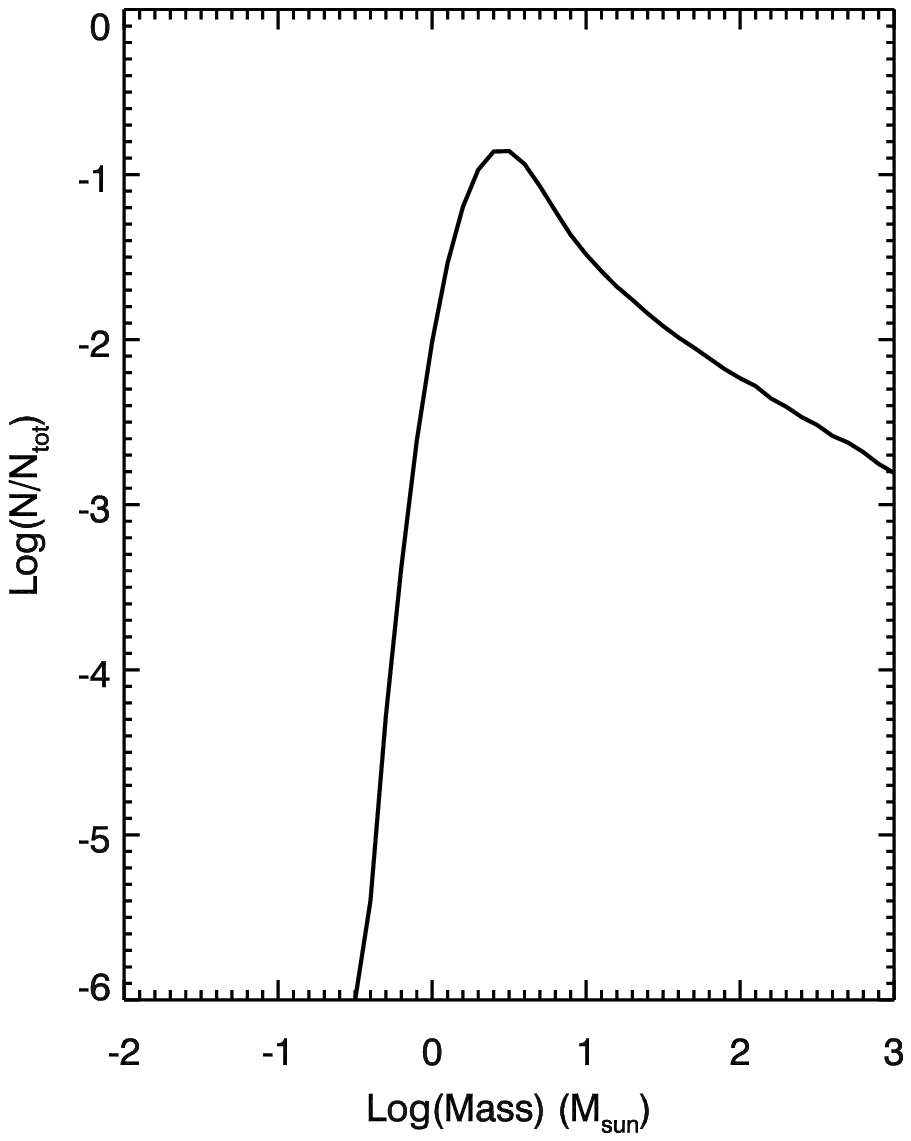}
\includegraphics[width=0.22\textwidth,angle=0]{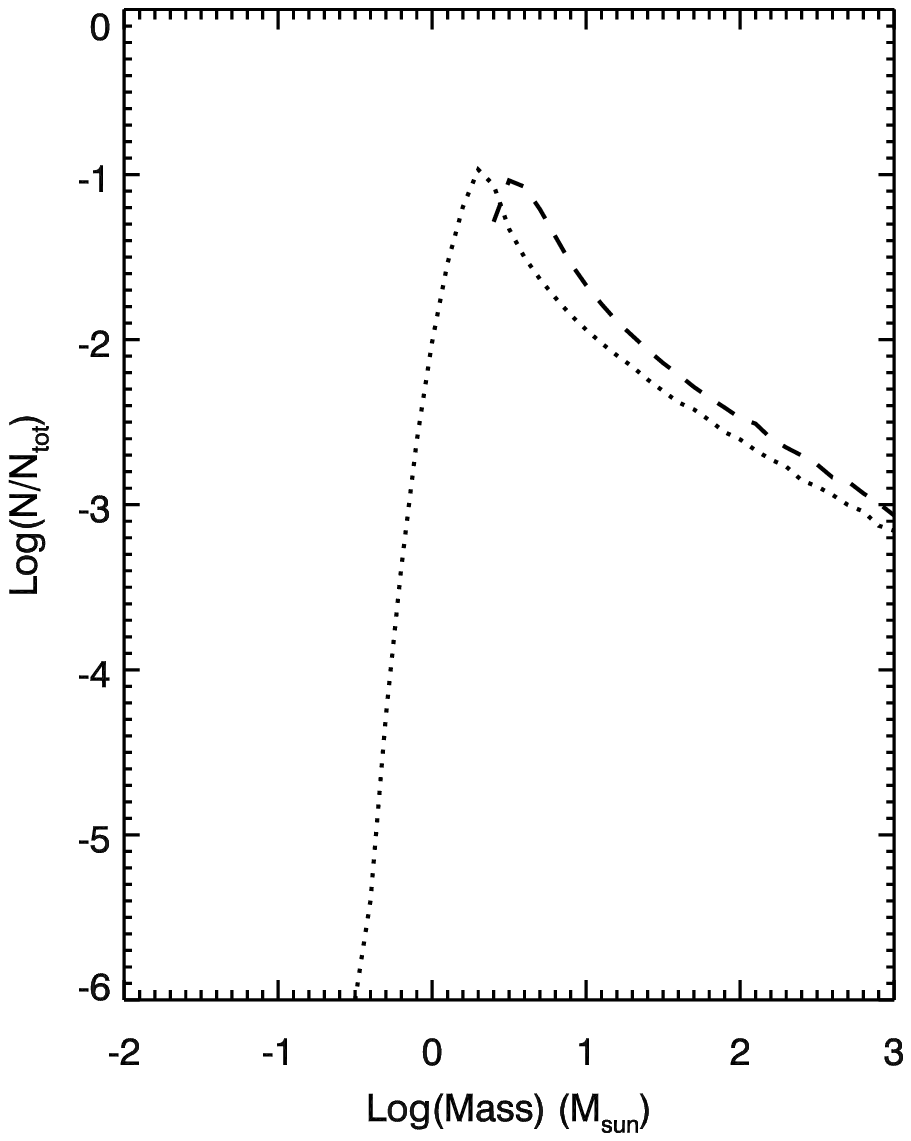}
\caption{Synthetic core mass functions for a flux frozen magnetic cloud assuming a broad, lognormal mass-to-flux ratio (FF2). Panels depict the same curves as Figure~\ref{cmfFF1}.}
\label{cmfFF2}
\end{figure}
\end{center}

\begin{center}
\begin{figure}
\centering
\includegraphics[width=0.22\textwidth,angle=0]{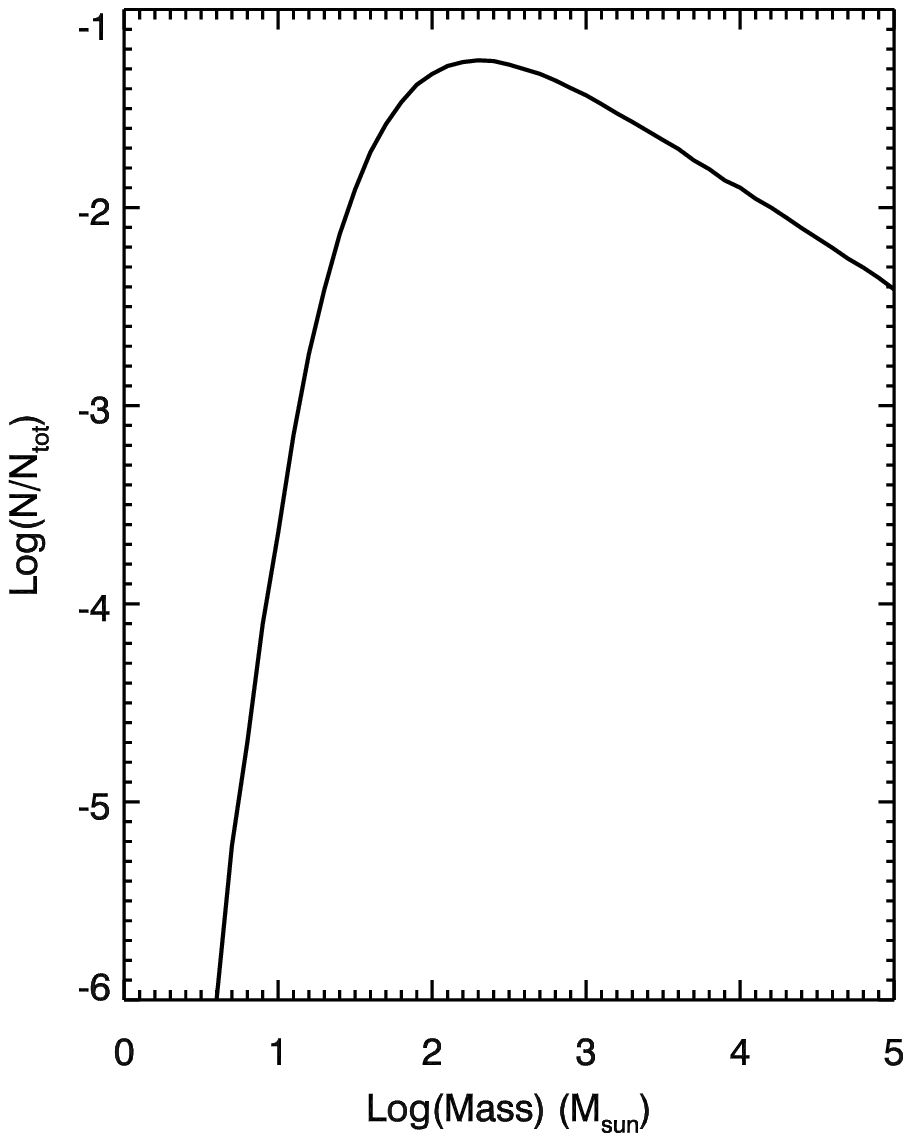}
\includegraphics[width=0.22\textwidth,angle=0]{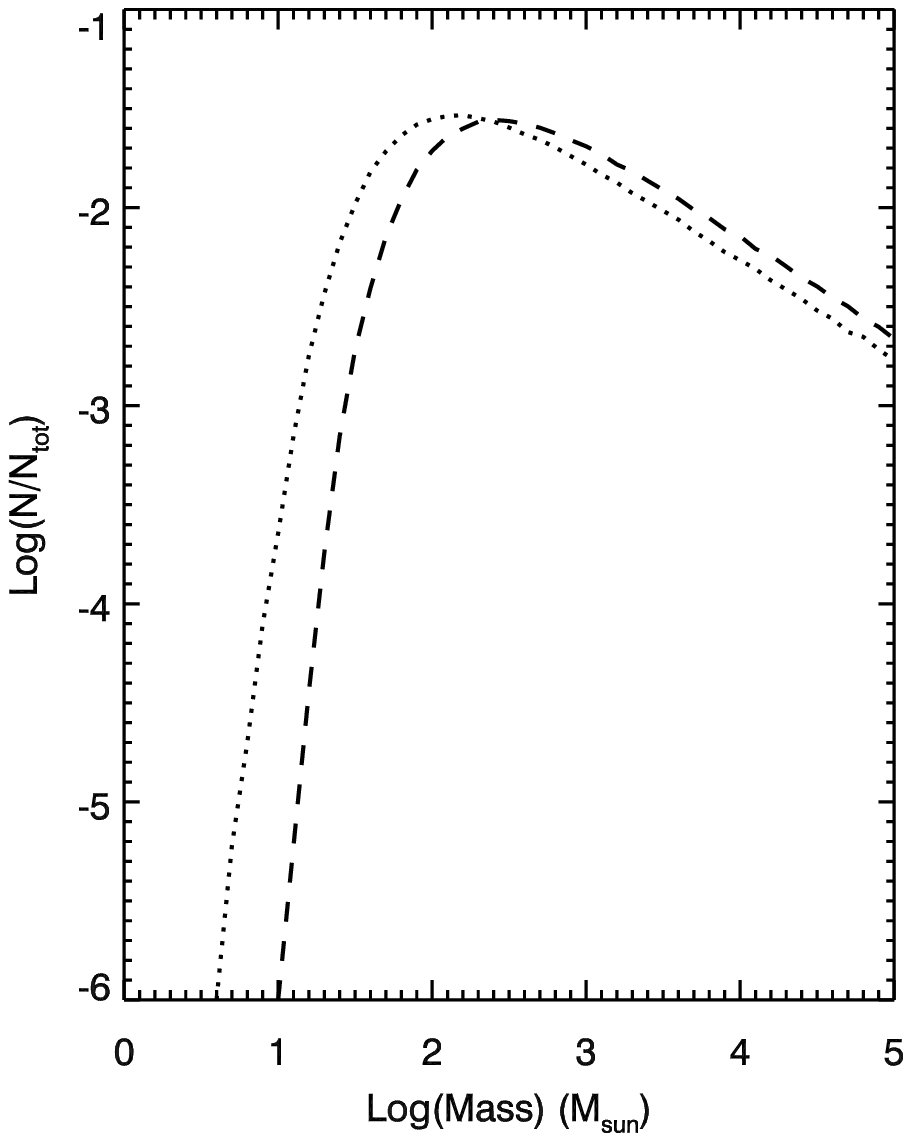}
\caption{Synthetic core mass functions for a flux frozen magnetic cloud assuming a narrow, lognormal mass-to-flux ratio (FF3). Panels depict the same curves as Figure~\ref{cmfFF1}.}
\label{cmfFF3}
\end{figure}
\end{center}

A main aim of this paper is to show the effect of a magnetic field on the CMF. A flux frozen field represents the simplest case. Such a scenario arises in highly ionized regions where frequent collisions between ions and neutral particles would ensure perfect coupling to the magnetic field. Figures~\ref{cmfFF1}-\ref{cmfFF3} show the resulting synthetic core mass function for the three models FF1, FF2, and FF3 respectively. Under the assumption of a uniform mass-to-flux ratio distribution (FF1), the resultant CMF (Figure~\ref{cmfFF1}, left) exhibits a narrow peak with a distinct high mass tail. The right hand panel of Figure~\ref{cmfFF1} again shows the contributions to the CMF from the two column density regimes ($A_{v} < 8$ mag (dashed line) and $A_{v} > 8$ mag (dotted line)). This composite plot shows that like the NM case, and in line with the Jeans theory, the low density gas forms high mass cores, while high density gas forms
low mass cores. However, unlike the Jeans theory and nonmagnetic case, we see that with the addition of a magnetic field, \textit{the high density gas also contributes to the formation of high mass cores}, albeit to a lesser extent. Compared to the NM case, the peak of this core mass function is shifted to $M \simeq 10^{0.7}\rm~M_{\odot} \simeq 5.0 \rm~M_{\odot}$. On the right hand side of this peak, the trend can be described by two distinct slopes. For  $0.7 < \log(M/M_{\odot}) < 1.2$, $\alpha = 0.8$ while for $\log(M/M_{\odot}) > 1.2$ the slope becomes shallower; $\alpha \sim 0.6$. Neither of these values correspond to the typical Salpeter and observational values. This discrepancy will be discussed further in Section~\ref{obs}.

The formation of the high mass tail is due to the relationship between $\mu_{0}$ and $\lambda$ as defined by Equation~\ref{lammu}. For $\mu_{0} - \sigma_{n}$ pairs which have mass-to-flux ratios closer to the critical value ($\mu_{0} = 1$), the corresponding length is up to 23 times larger than the thermal Jeans length for the same column density (see Figure~\ref{lammufig}). This increase in length scale has a direct effect on the mass of the core that is formed. Conversely, the low mass distribution is formed by $\mu_{0} - \sigma_{n}$ pairs that have low column density and mass-to-flux values that are closer to the other limit ($\mu_{0} = 3$), where $\lambda$ is only about 1.5 times larger than the thermal length scale.     

Figure~\ref{cmfFF2} shows the resulting synCMF for a broad lognormal $\mu_{0}$ distribution (FF2). The two panels again show the total and composite CMFs as described above. 
\begin{center}
\begin{figure}
\centering
\includegraphics[width=0.22\textwidth,angle=0]{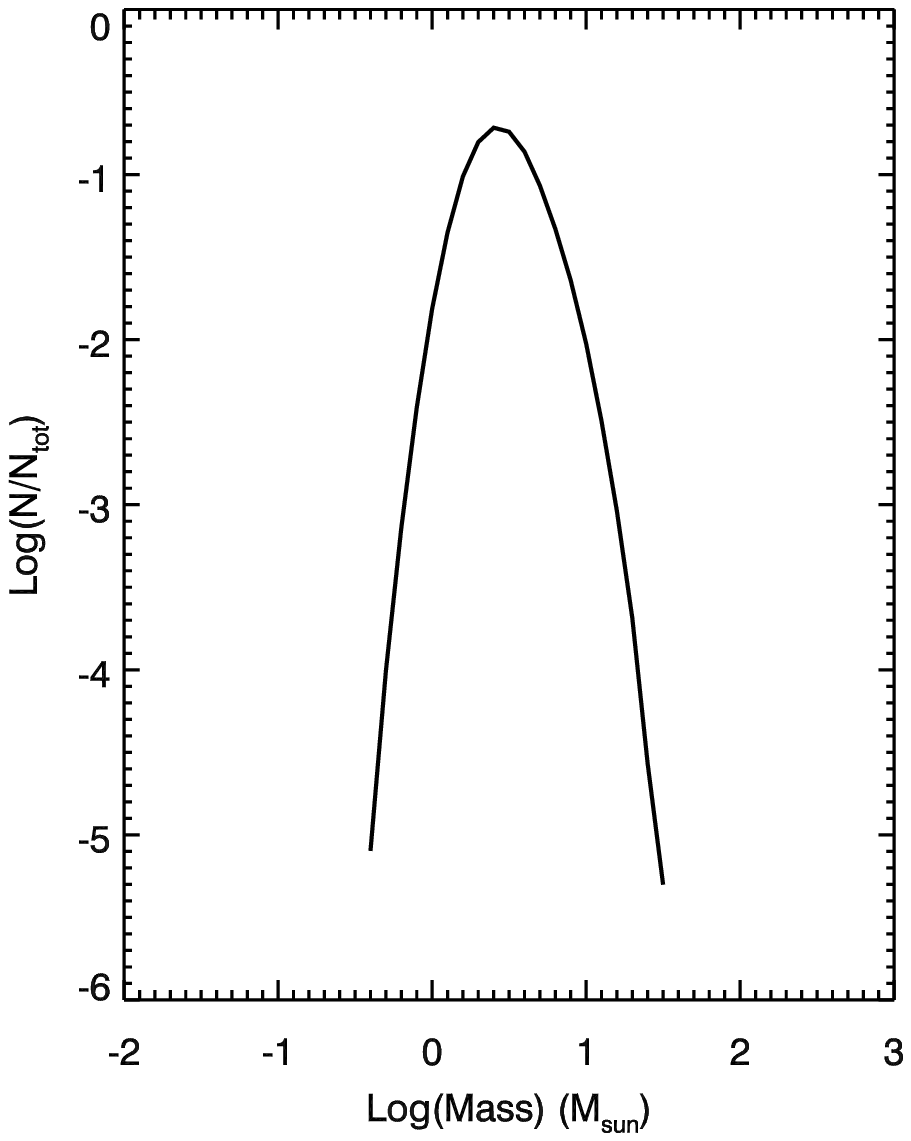}
\includegraphics[width=0.22\textwidth,angle=0]{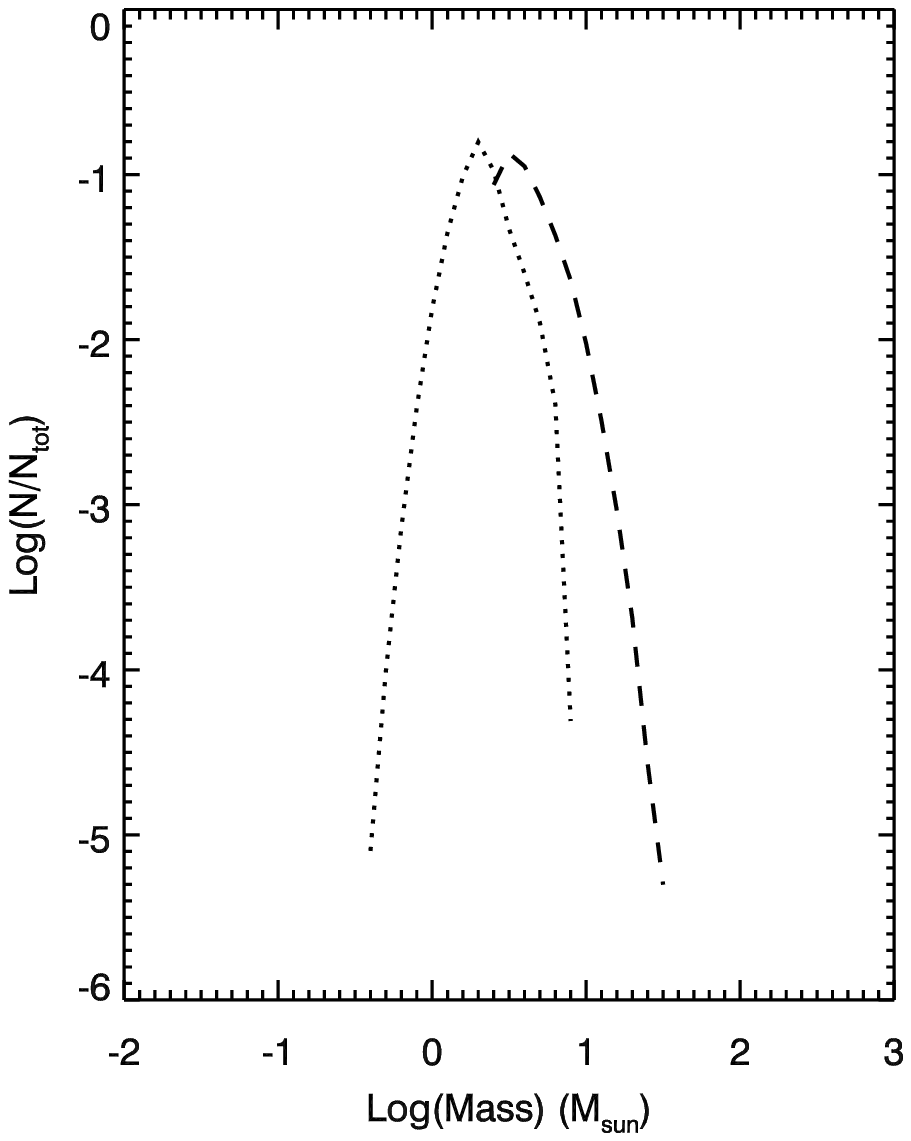}
\caption{Synthetic core mass functions for a magnetic cloud including the effects of ambipolar diffusion assuming a uniform subcritical distributed mass-to-flux ratio (AD1). Panels depict the same curves as Figure~\ref{cmfFF1}.}
\label{cmfAD5}
\end{figure}
\end{center}
\begin{center}
\begin{figure}
\centering
\includegraphics[width=0.22\textwidth,angle=0]{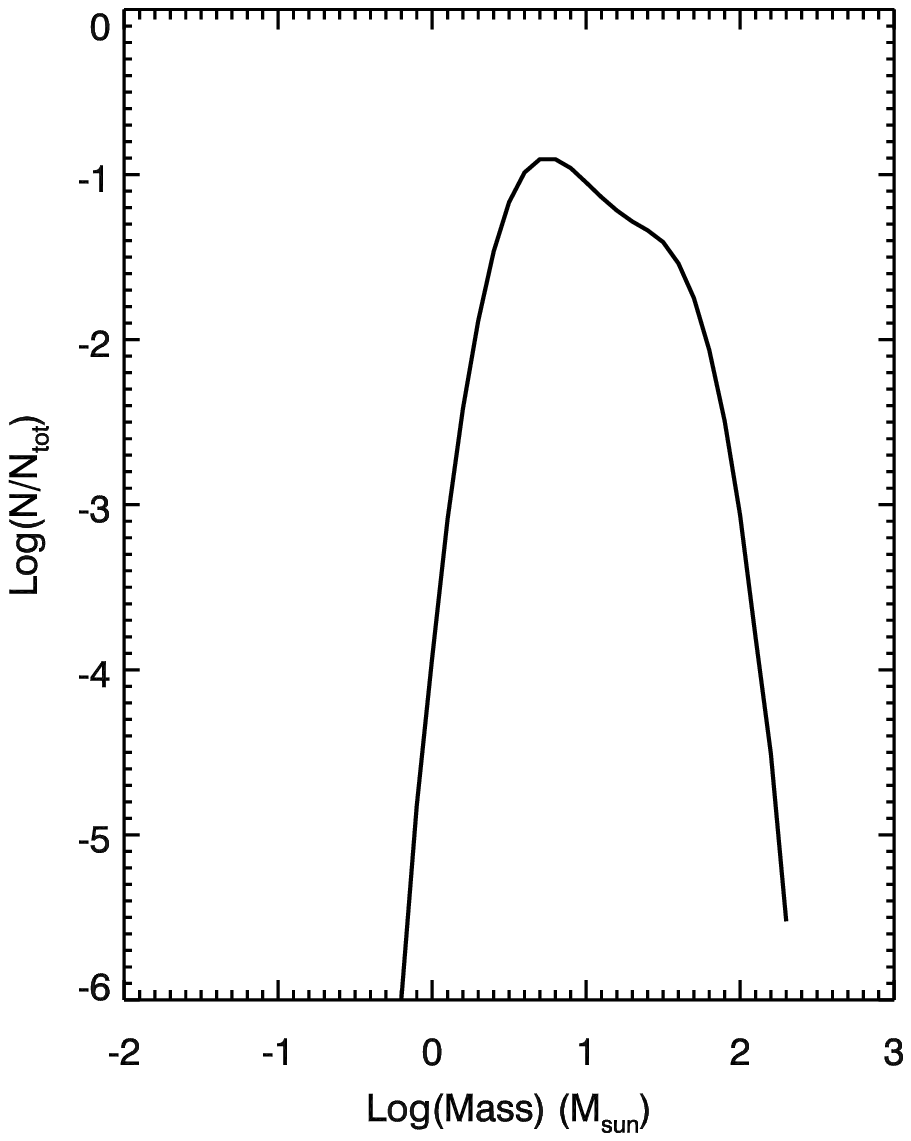}
\includegraphics[width=0.22\textwidth,angle=0]{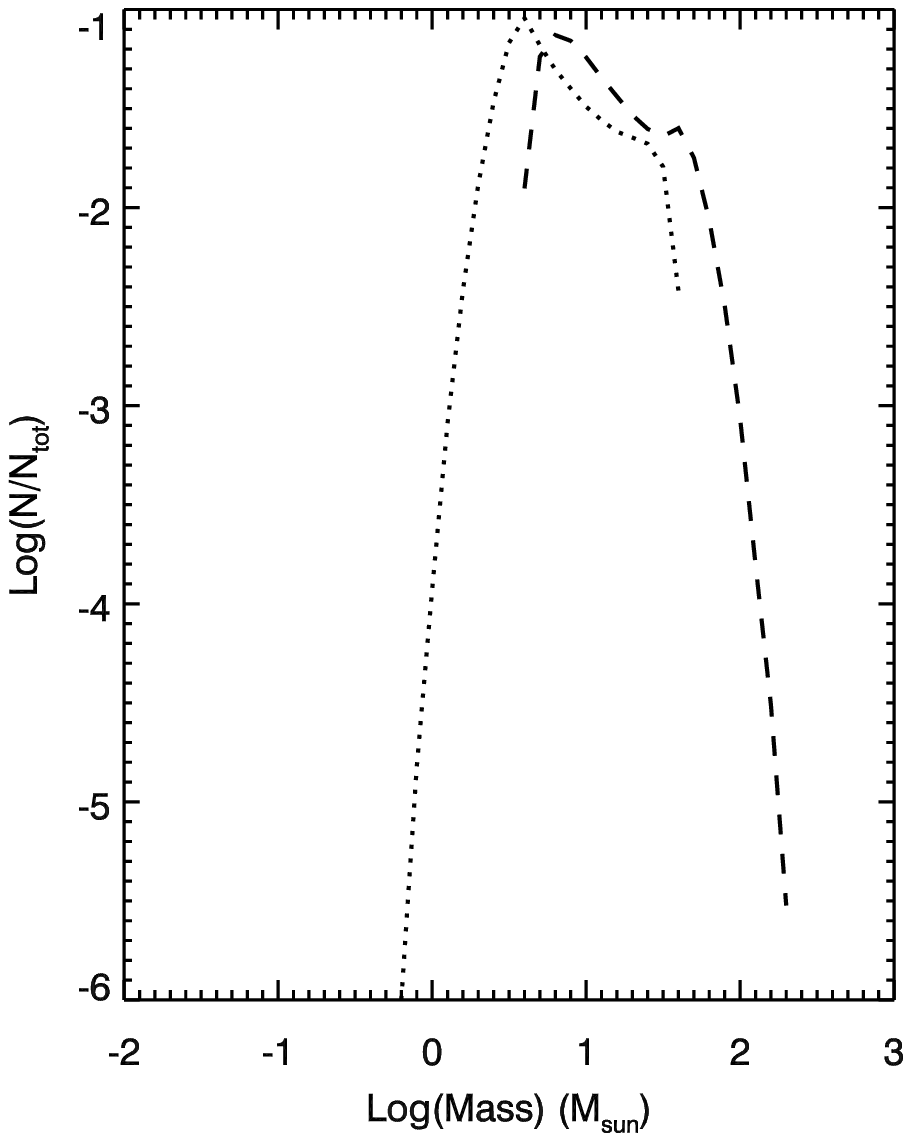}
\caption{Synthetic core mass functions for a magnetic cloud including the effects of ambipolar diffusion assuming a uniform supercritical distributed mass-to-flux ratio (AD2). Panels depict the same curves as Figure~\ref{cmfFF1}.}
\label{cmfAD2}
\end{figure}
\end{center}
This distribution results in a CMF that is similar to that of model FF1 (Figure~\ref{cmfFF1}), with a few minor differences. First, the high mass tail exhibits a steeper slope that results in a more pronounced peak region. Second, the peak of the mass function has shifted to a slightly smaller value of $M \simeq 10^{0.5} \rm{~M_{\odot}} = 3.16 \rm ~M_{\odot}$. As before, the trend of the high mass side can be described by two distinct slopes. For  $0.5 < \log(M/M_{\odot}) < 1.0$, $\alpha = 1.31$ while for $\log(M/M_{\odot}) > 1.0$ the slope becomes shallower; $\alpha = 0.63$.

Figure~\ref{cmfFF3} shows the resulting synCMF for a narrow log-normal $\mu_{0}$ distribution (FF3). Unlike the previous two models, this one does not exhibit a narrow log-normal type peak, but rather shows a broad peak that leads directly into a high mass tail. As a result, the post peak trend for this model can be described by a single slope, $\alpha = 0.44$. Also, note that the function itself has been shifted toward higher masses as compared to the other two flux frozen models. As such, this CMF peaks at $M \simeq 10^{1.3} \rm{~M_{\odot}} \sim 20~\rm M_{\odot}$. This shift in the mass range is due entirely to the narrow peak distribution of the mass-to-flux ratio; all of the chosen mass-to-flux ratios result in length scales that are $\sim~6-23$ times larger than the thermal length scale (see Figure~\ref{lammufig})  and therefore, the low mass cores that are formed in the other two models are absent in this model. Overall, as shown by all three 
\begin{center}
\begin{figure*}
\centering
\includegraphics[width=0.22\textwidth,angle=0]{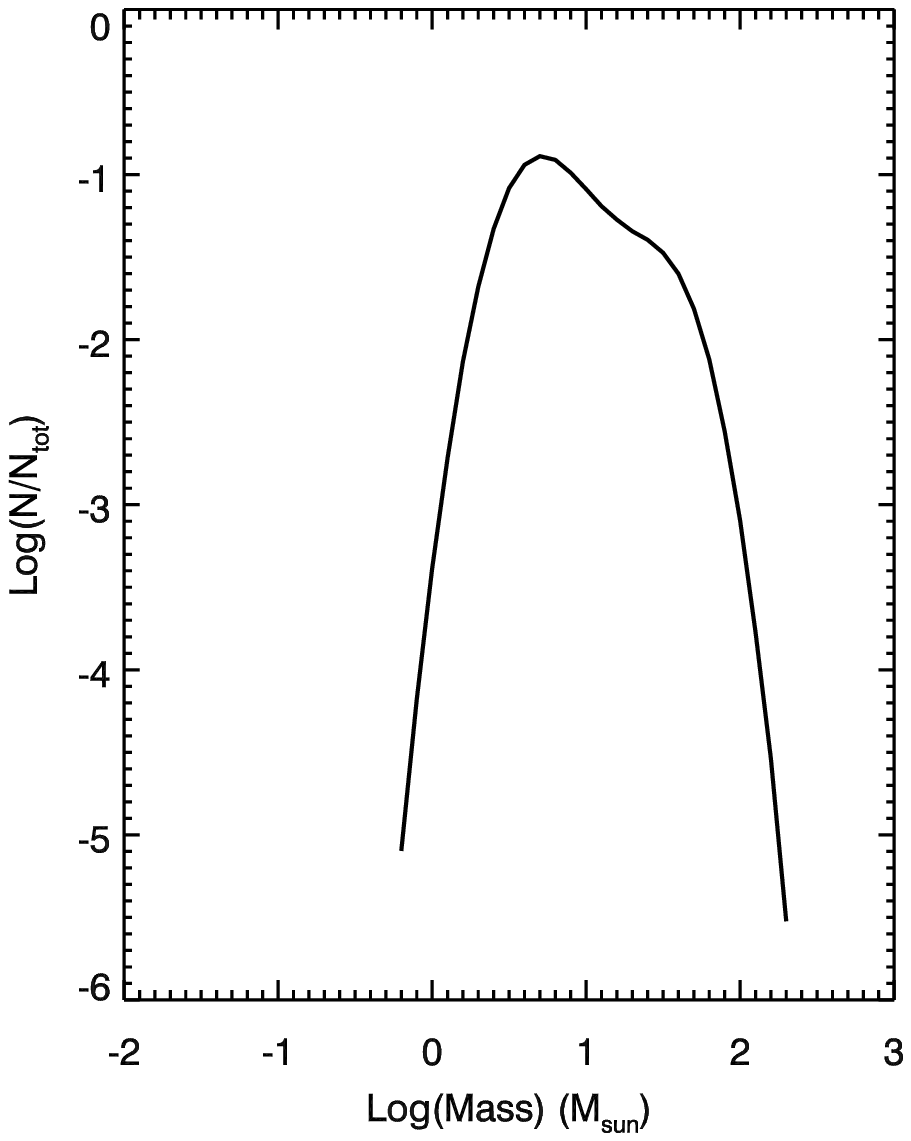}
\includegraphics[width=0.22\textwidth,angle=0]{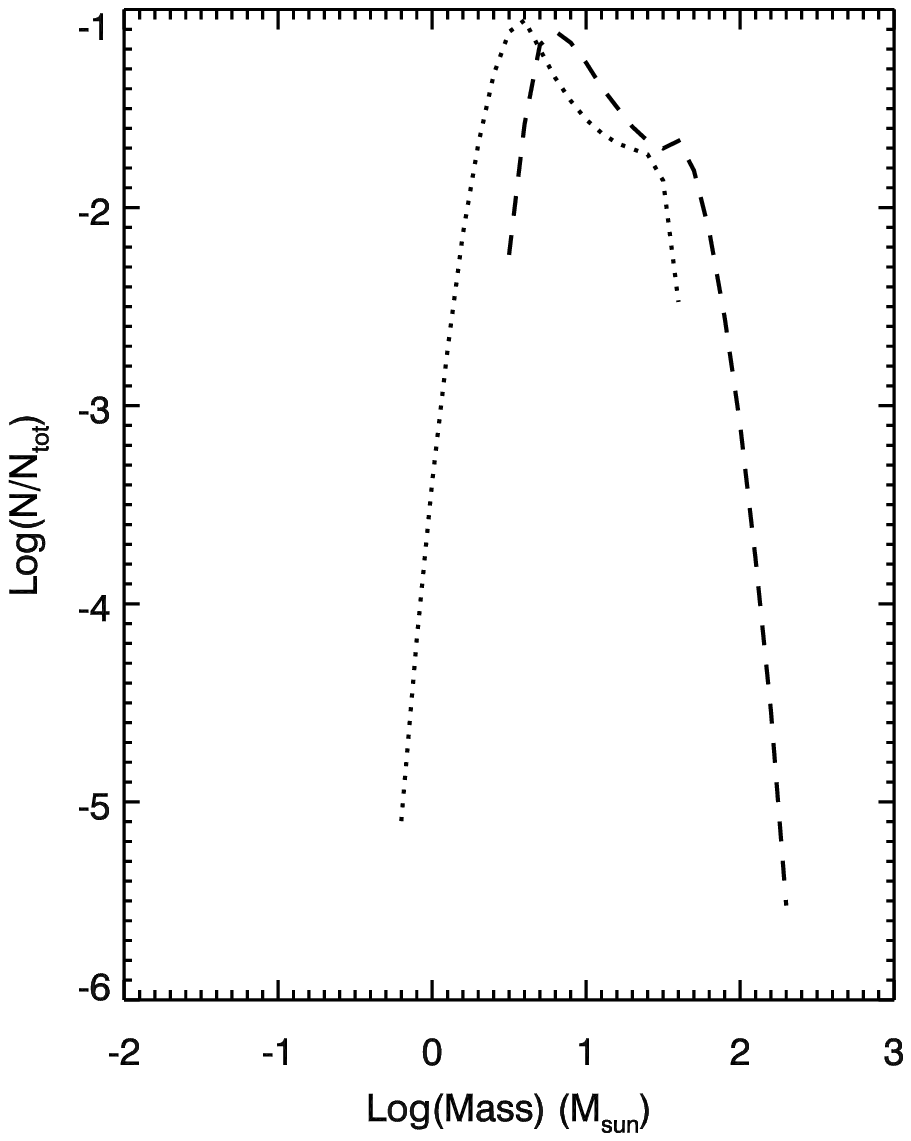}
\includegraphics[width=0.22\textwidth,angle=0]{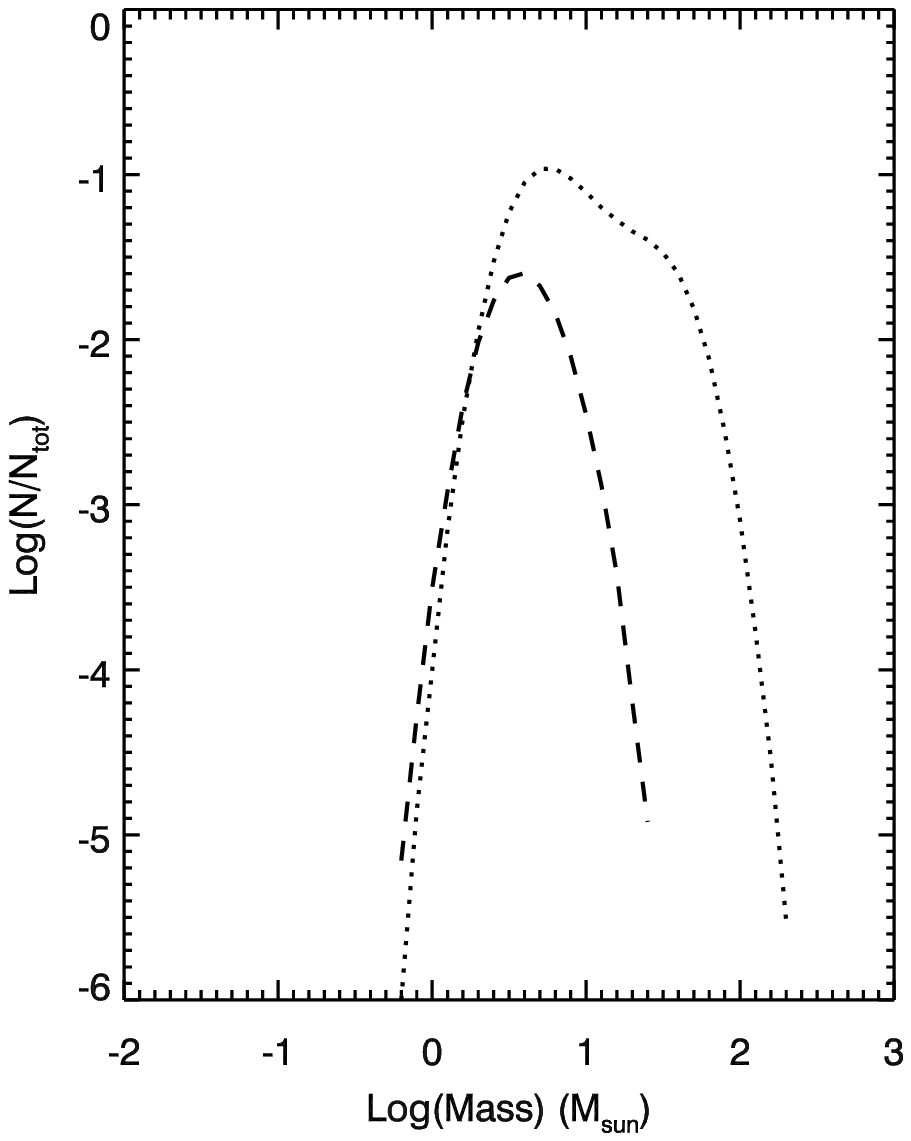}
\caption{Synthetic core mass functions for a magnetic cloud including the effects of ambipolar diffusion assuming a uniformly distributed mass-to-flux ratio (AD3). Left: Total core mass function. Middle: Contributions to the core mass function from cores with $A_{v} < 8$ mag (dashed line) and cores with $A_{v} > 8$ mag (dotted line). Right: Contributions to the core mass function from cores with $\mu_{0} < 1$ (dashed line) and cores with $\mu_{0} > 1$ (dotted line). 
}
\label{cmfAD1}
\end{figure*}
\end{center}
models, the effect of adding a flux-frozen field is the appearance of a broad shallow tail at the high mass end of the core mass function. 

\begin{center}
\begin{figure*}
\centering
\includegraphics[width=0.22\textwidth,angle=0]{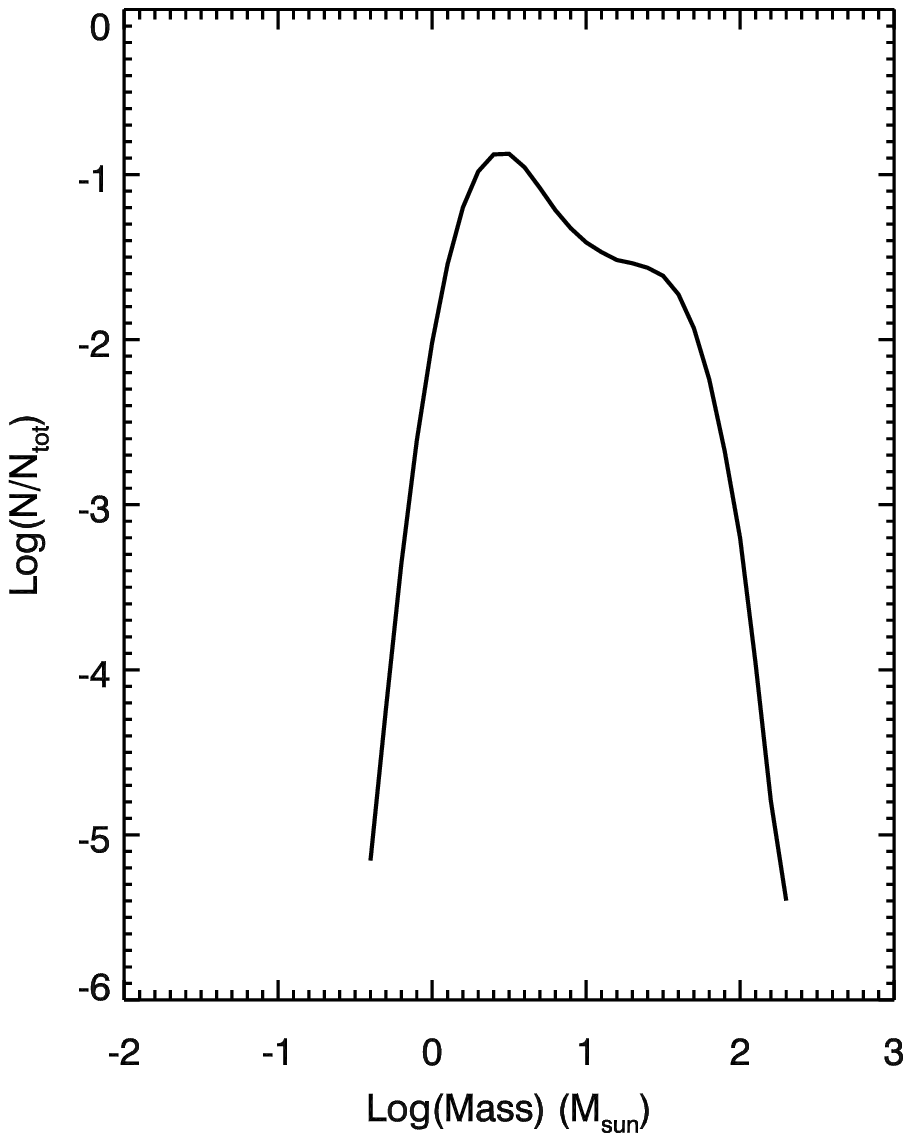}
\includegraphics[width=0.22\textwidth,angle=0]{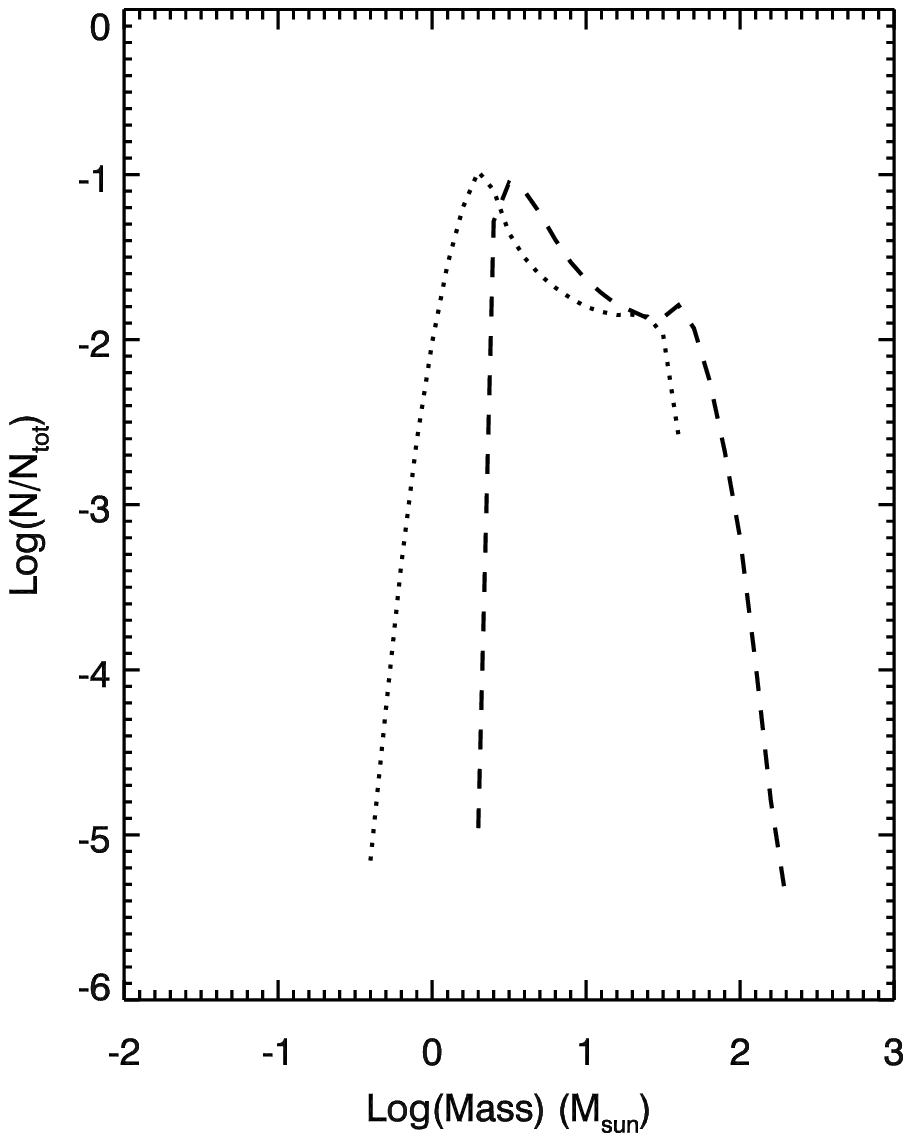}
\includegraphics[width=0.22\textwidth,angle=0]{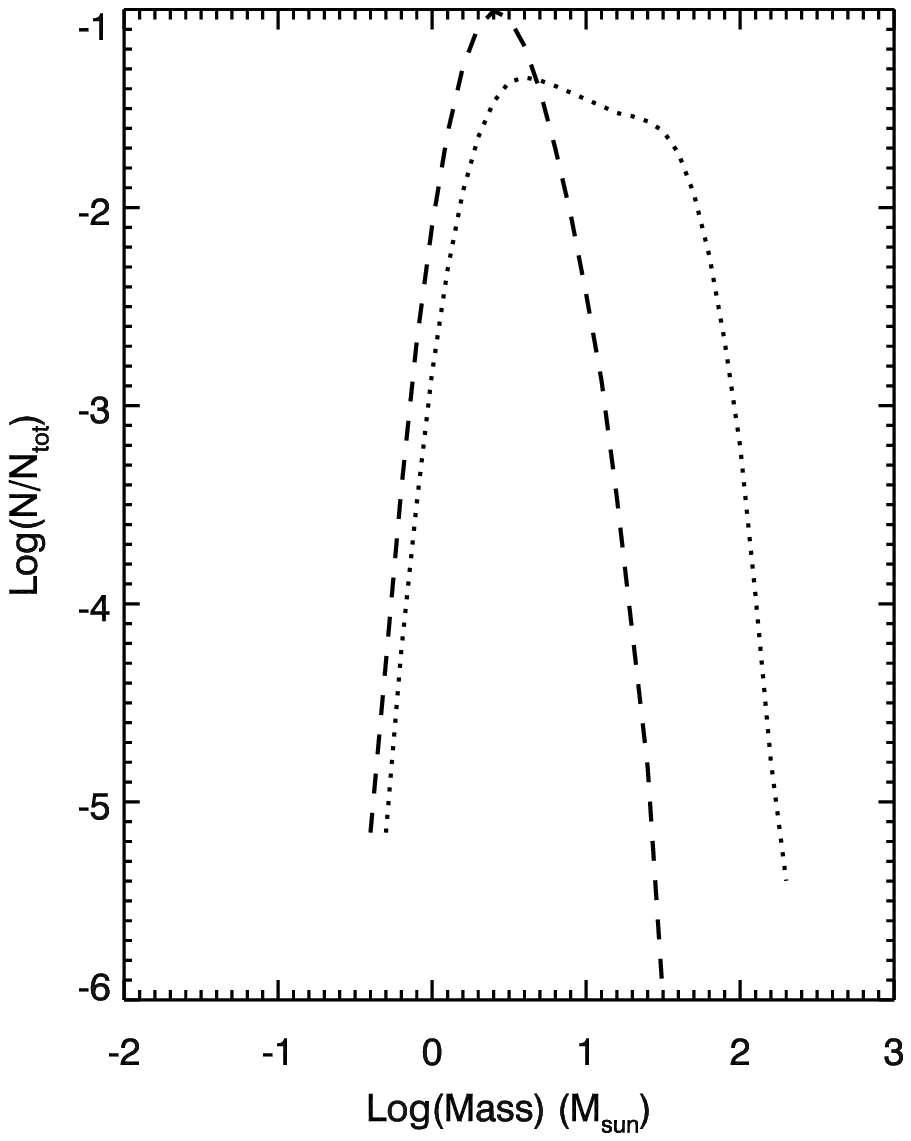}
\caption{Synthetic core mass functions for a magnetic cloud including the effects of ambipolar diffusion assuming a broad, lognormal mass-to-flux ratio distribution (AD4). Panels depict the same curves as Figure~\ref{cmfAD1}.}
\label{cmfAD3}
\end{figure*}
\end{center}
\begin{center}
\begin{figure*}
\centering
\includegraphics[width=0.22\textwidth,angle=0]{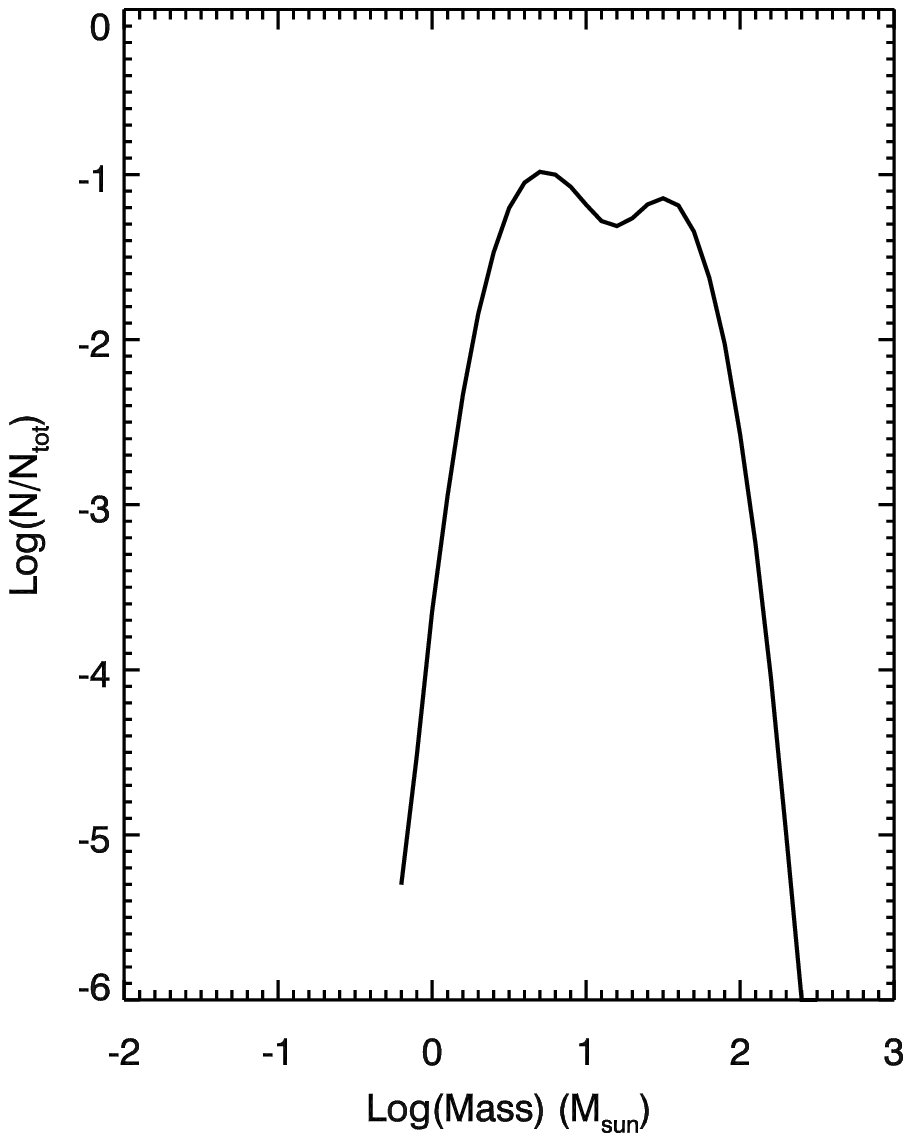}
\includegraphics[width=0.22\textwidth,angle=0]{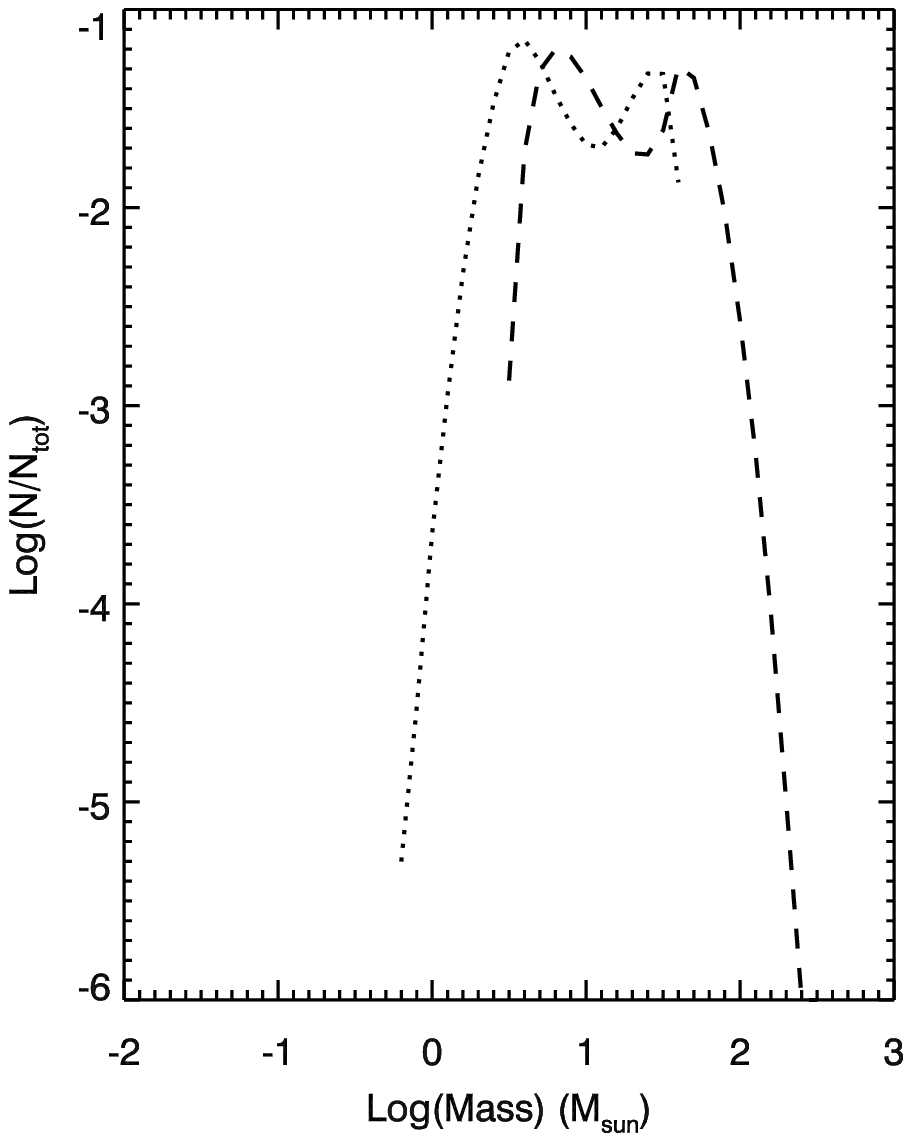}
\includegraphics[width=0.22\textwidth,angle=0]{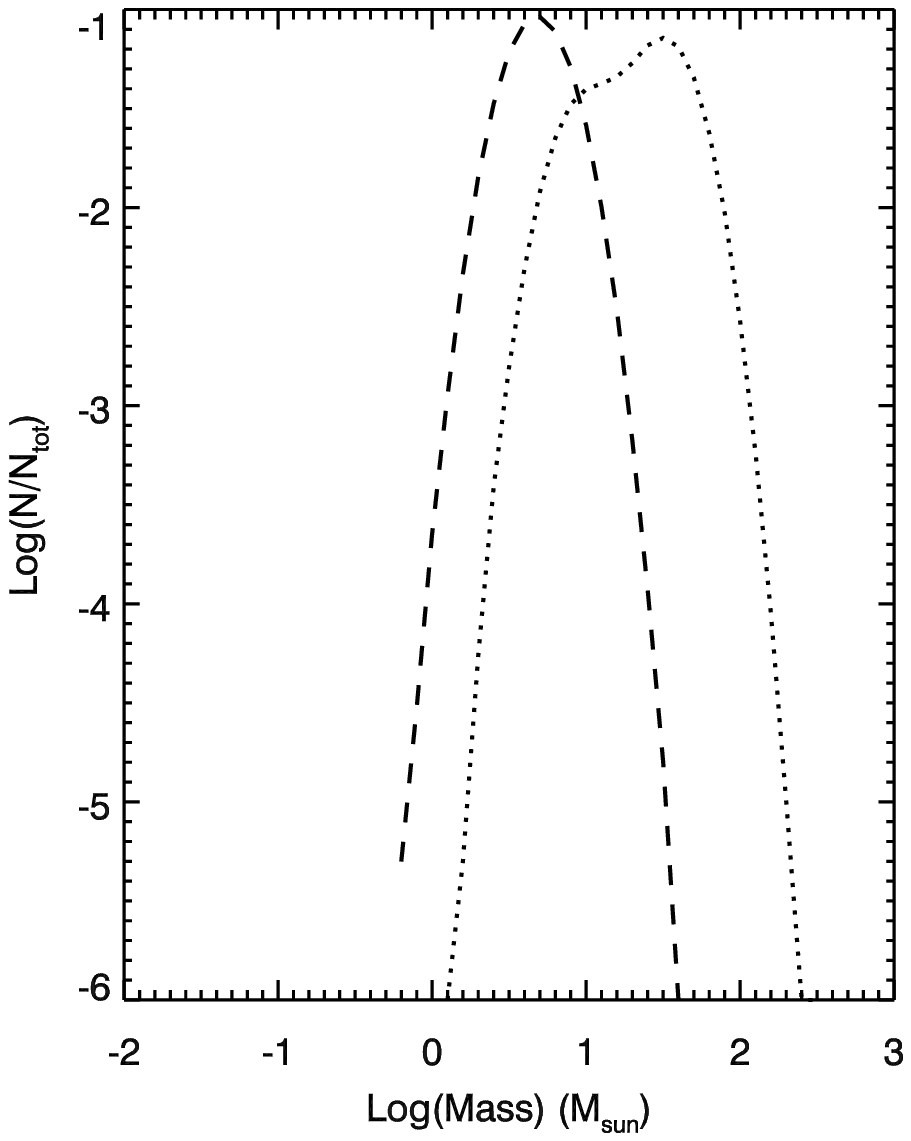}
\caption{Synthetic core mass functions for a magnetic cloud including the effects of ambipolar diffusion assuming a narrow, lognormal mass-to-flux ratio distribution (AD5). Panels depict the same curves as Figure~\ref{cmfAD1}.}
\label{cmfAD4}
\end{figure*}
\end{center} 

\subsection{Ambipolar Diffusion Magnetic Model}

In the previous section we looked at the effect of a simple flux-frozen field on the shapes of the resulting CMF(s). Here we look at how the addition of neutral-ion slip via ambipolar diffusion affects the shape of the CMF. As discussed above, we have fixed the normalized neutral-ion collision time to $\tau_{ni,0}/t_{0} = 0.2$. This implies a high degree of ambipolar diffusion and therefore less frequent collisions between the neutrals and ions. Such a situation would occur in the inner regions of a molecular cloud where the main ionization mechanism is cosmic rays.

Figures~\ref{cmfAD5}-\ref{cmfAD4} show the resulting synCMFs for all five mass-to-flux ratio distributions respectively. To establish how the sub- and supercritical regions of the mass-to-flux ratio affect the shape of the CMF, we start our analysis by presenting two cases that isolate each regime. Figures~\ref{cmfAD5}~\&~\ref{cmfAD2} show the resulting synCMFs for the subcritical and supercritical uniform mass-to-flux ratio distributions (AD1 and AD2) respectively. The two panels show the total and constituent core mass functions as described in the previous section.  

Focusing on model AD1, Figure~\ref{cmfAD5}, the left panel shows that the core mass function is very similar to the nonmagnetic model (see Figure~\ref{cmfNM}, left). This is due to the fact that the curve on the subcritical side of Figure~\ref{lammufig} converges to the nonmagnetic limit faster than in the trans- and supercritical regions. Upon closer comparison, AD1 peaks at the approximately the same value as NM, however the density composite CMF (Figure~\ref{cmfAD5}, right) reveals differences between these two models. Unlike the nonmagnetic model, AD1 shows evidence that a portion of the high column density gas goes toward forming high mass cores (Figure~\ref{cmfAD5}, right).  

Figure~\ref{cmfAD2} shows the resulting synCMF under the assumption of a uniform supercritical distribution (AD2). The left panel shows that the total CMF is a hybrid between the nonmagnetic and flux-frozen models presented above. Specifically, this CMF shows the same peaked nature with high mass tail as the flux frozen model, however this tail abruptly declines at about 100 M$_{\odot}$. This truncation makes the over all shape of the CMF resemble the nonmagnetic case, albeit broader, with the beginnings of a ``shoulder'' feature between 10 and 100 M$_{\odot}$. Looking at the composite column density CMF (Figure~\ref{cmfAD2}, right), we see that the lowest and highest mass cores are formed by the highest and lowest density gas respectively, while the middle has contributions from both density regimes. The peak of the mass function for this model occurs at about $\log(M/M_{\odot}) = 0.7$.  

Model AD3 assumes a uniform mass-to-flux ratio distribution that samples the peak of the $\lambda$ versus $\mu_{0}$ graph (see Figure~\ref{lammufig}). The resulting CMF (Figure~\ref{cmfAD1}, left) is very similar to the one produced by AD2. Looking at the contributions from the low and high column density gas, we again see that the lowest and highest mass cores are formed by the highest and lowest density gas respectively while the middle range has contributions from both density regimes. 

The right panel of Figure~\ref{cmfAD1} shows the contributions from the subcritical ($\mu_{0}\le 1$, dashed line) and supercritical ($\mu_{0} > 1$,  dotted line) gas. We see that the total synCMF for AD3 (Figure~\ref{cmfAD1}, left) is a combination of models AD1 and AD2. Specifically, we see that the majority of the cores are formed from supercritical gas, while the subcritical gas yields a minor contribution to the population of low mass cores. By mentally combining the middle and right hand plots in Figure~\ref{cmfAD1}, one can determine that the highest mass cores are formed by supercritical gas and fall into the non-star-forming regime while low-mass cores are formed by both supercritical and subcritical gas, and fall into both the star-forming and non-star-forming regimes. The peak of the mass function for this model occurs at about $\log(M/M_{\odot}) = 0.7$ and the average slope of the high mass `tail' is $\alpha = 1.42$.

Finally, Figures~\ref{cmfAD3}~\&~\ref{cmfAD4} show the resulting synCMFs for the two lognormal $\mu_{0}$ distributions, AD4 and AD5, respectively. The broad lognormal distribution (AD4) is similar to models AD2 and AD3, however this model shows a more distinct `peak' and `shoulder' region as compared to the other two. Looking at the composite mass-to-flux ratio plot (Figure~\ref{cmfAD3}, right) we see that the peak region is mainly formed by subcritical gas while the shoulder region is formed mainly by contributions from supercritical gas. This model peaks at $M = 10^{0.7} ~M/M_{\odot} \simeq 5.0~\rm M_{\odot}$, and the average slope of the high mass tail is $\alpha = 1.18$. Switching to the narrow lognormal distribution (Figure~\ref{cmfAD4}), we see that this model results in a double peaked function. Examination of the composite plots show that the low mass peak is formed by the subcritical material while the second peak is formed by supercritical material. These peaks occur at $\log(M/M_{\odot}) \sim 0.7$ and $\log(M/M_{\odot}) \sim 1.5$ respectively. The formation of the high mass peak is due to the extremely narrow mass-to-flux ratio distribution. It picks out only large length scales from the peak of the $\lambda - \mu_{0}$ curve (with $\tau_{ni,0}/t_{0} = 0.2$) in Figure~\ref{lammufig}.    

\begin{center}
\begin{figure*}
\centering
\includegraphics[scale=0.8,angle=0]{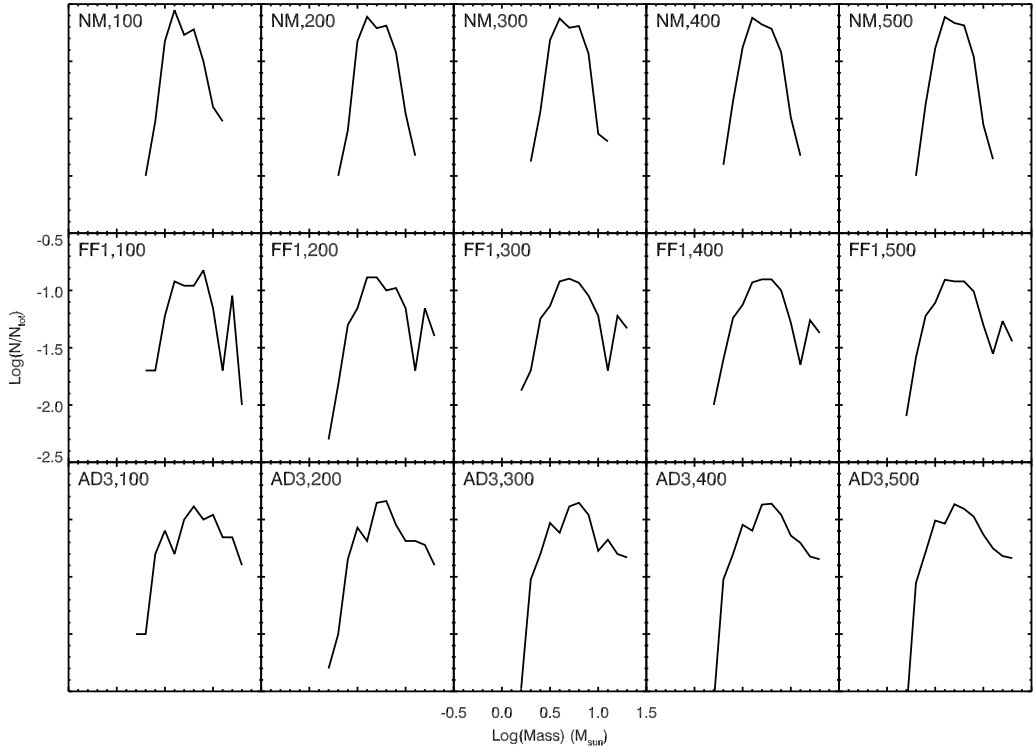}
\caption{Small sample core mass functions for three models: NM (top row), FF1 (middle row), and AD1 (bottom row). Number of cores for each panel indicated in the top left hand corner of each plot.}
\label{smallsample}
\end{figure*}
\end{center}

\subsection{Assessment of Synthetic Core Mass Functions}
\label{assess}
The previous subsections presented the overall results and features of each of the models. Within these results we found three main features that changed between the different models. These are the overall shape of the core mass function, the location of the peak(s) and the slope of the high mass tail (if it exists). Here we discuss these three features across all models. 

\begin{center}
\begin{figure*}
\centering
\includegraphics[scale=0.8,angle=0]{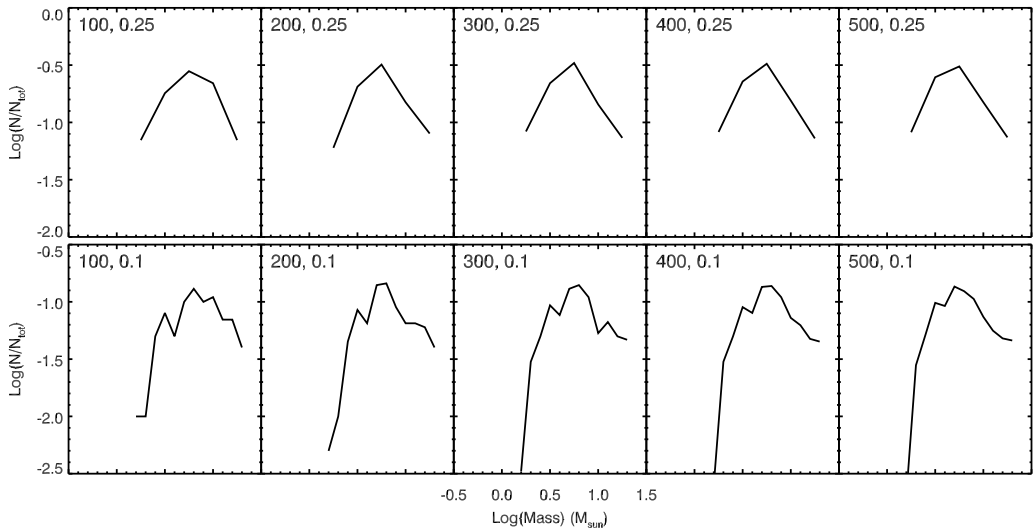}
\caption{Bin size comparison for small sample core mass functions. Panels show the effect of the bin size on the resulting curve for  $\Delta\log(M/M_{\odot}) =0.25$ bins (top row) and $\Delta\log(M/M_{\odot}) = 0.1$ bins (bottom row). Model used in all panels is AD3.}
\label{binsize}
\end{figure*}
\end{center}

\subsubsection{Shape}
Within the nine models presented, there were three distinct recurring shapes; pure lognormal as represented by the NM and AD1 models, lognormal peak with a shoulder as represented by AD2, AD3, AD4 and AD5, and the lognormal peak with high mass tail as represented by FF1, FF2, and FF3. The appearance of these shapes are directly connected to the state of the magnetic field in the region. In the absence of a magnetic field, the CMF is a pure lognormal function. This shape is also observed in model AD1. As mentioned earlier, the reason that this AD model shows such a shape while the other ones do not is due to the shape of the $\lambda - \mu_{0}$ curve on the subcritical side of Figure~\ref{lammufig}; the curve asymptotes to the nonmagnetic limit faster on that side than on the supercritical side. Therefore one would expect a model with only subcritical mass-to-flux values to look similar to the nonmagnetic model, but with a slight broadening due to a narrow region of mass-to-flux ratios with $\lambda$ larger than the non-magnetic limit. 

For models with an increasing supercritical regime, the broadening becomes more pronounced as a shoulder develops. This shoulder is due to an increase in higher mass cores that are the product of the larger length scales picked out by the supercritical mass-to-flux ratios. The extent of the shoulder depends on the mass-to-flux ratio distributions. For uniform distributions, the CMF is narrower with a less defined shoulder region, while for a broad lognormal distribution, the shoulder region is much broader and distinct. Finally, the appearance of the double peaked CMF in AD5 is an example of an extreme shoulder. This second peak is due solely to the extremely narrow mass-to-flux ratio range used in this model. This preferentially picks out only mass-to-flux ratios with length scales much larger than the nonmagnetic model. 

The appearance of the pure high mass tail is entirely a product of flux-freezing. This is due to the asymptotic nature of the flux-frozen curve as it nears the critical mass-to-flux ratio (see Figure~\ref{lammufig}). This allows for transcritical mass-to-flux ratio values to produce much larger masses for the same column density.

\subsubsection{Peak Location}

The location of the CMF peak depends on the distribution of the mass-to-flux ratio. The location of the peak in the nonmagnetic case, which occurs at $\log(M/M_{\odot}) = 0.4$ ($M \sim 2.5$ M$_{\odot}$) serves as the comparison point. For magnetic models, the location of the peak was generally larger than this value as long as the mass-to-flux ratio distribution was uniform with some contribution from the supercritical regime (see models FF1, AD2, and AD3). Model AD1, although also assuming a uniform mass-to-flux ratio distribution, exhibits a similar peak value to NM due to the exclusion of supercritical mass-to-flux ratio values. When considering the lognormal mass-to-flux ratio distributions, we find that the peak location is dependent on the width of the distribution. Specifically, broader distributions exhibit values closer to the NM peak value, while narrower distributions exhibit peak values that are higher than the nonmagnetic case. Model AD5 is an anomaly and does not fit within these trends given that it exhibits two peaks. 

\subsubsection{High Mass Slope}

As alluded to earlier, the shape and extent of the high mass slope was found to be variable and connected to the influence of the magnetic field. Specifically, the appearance of the `shoulder' feature is directly connected to the presence of ambipolar diffusion. The degree of the shoulder in the ambipolar diffusion models was found to be dependent on the range of allowed mass-to-flux ratio values. Overall, these differences in shapes result in a wide range of slopes. For the flux-frozen models, the slopes were as steep as $\alpha = 1.31$ in the case of  FF2, and as shallow as $\alpha = 0.44$ in the case of FF3. For the ambipolar diffusion models, the average high mass slope ranges between $\alpha = 1.18$ and $\alpha = 1.42$. Although some of these slopes are consistent with the Salpeter value, $\alpha = 1.35$ \citep{Salpeter1955}, others are significantly different. Further analysis of this discrepancy  is given in the following section.  

\section{Scaling to Observations}
\label{obs}

Unlike our synCMFs, typical observational CMFs usually contain on the order of 200 cores, not 10$^{6}$. Therefore, to make our analysis relevant for typical observed CMFs, we must scale back our sample sizes to those typically observed. The following two sections explore the effect of two observational constraints, sample size and bin size, on the shape and slope of observed CMFs.

\subsection{Effect of Sample Size}

To test the effect of the sample size on the resultant CMF, we scaled three synCMFs (NM, FF1, and AD3) down to plausible observational sample sizes (100, 200, 300, 400, and 500 cores). Figure~\ref{smallsample} shows the resulting synCMFs for each of the fifteen cases. In addition to scaling the sample size, we have also truncated the mass range considered to one more typically found in observed CMFs ($-1.0 < \log(M/M_{\odot}) < 1.3$). Under these scaled conditions, we see that the nonmagnetic CMFs still maintain the overall shape exhibited by the full sample curve (Figure~\ref{cmfNM}), however the two magnetic cases are fairly different. The high mass tail and truncated shoulder features present in the full sample curves for FF1 and AD3 respectively are no longer quite as distinct at these sample sizes. For a definitive difference between the ambipolar diffusion and flux-frozen cases, observations would have to extend up to objects with masses between 10$^{2}$ and 10$^{3}$ solar masses. Therefore, on typical observational scales, conclusions about the nature of the magnetic field from the shape of the CMF are possible, but highly uncertain.

\subsection{Effect of Bin Size}

Constructing histograms for the purposes of determining a CMF requires binning data into predetermined mass bins. For the above synCMFs, we used $\Delta\log(M/M_{\odot}) = 0.1$ size bins. Variations in the bin size acts to change the resolution of the resulting curve; smaller bins yield more detail while larger bins show only the broad strokes. To determine the effect of the bin size on the resulting CMF, we re-binned the histograms for AD3 in Figure~\ref{smallsample} (bottom row) using $\Delta\log(M/M_{\odot}) =0.25$ bins. Figure~\ref{binsize} shows the comparison of the original bin size ($\Delta\log(M/M_{\odot}) = 0.1$) to the new bin size (top row). As expected, with the larger bin size, the detail becomes smeared out, resulting in an average curve.

\begin{center}
\begin{figure}
\includegraphics[scale=0.33,angle=-90]{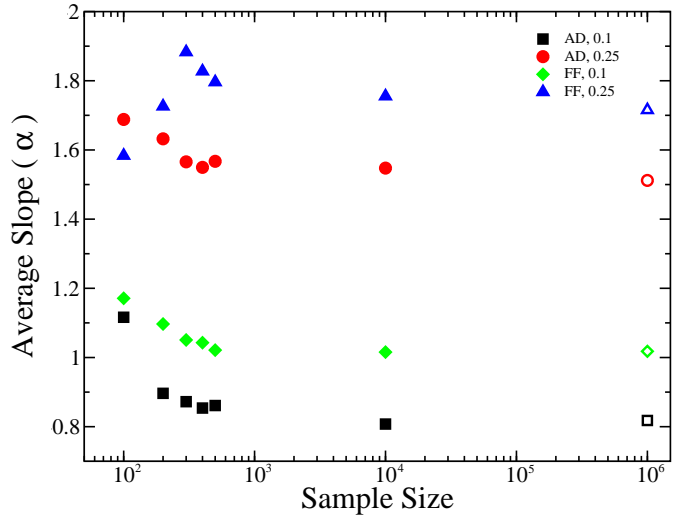}
\caption{Average slope as a function of sample size. Symbols represent the derived slopes for the two models and two bin sizes: AD3, $\Delta\log(M/M_{\odot}) = 0.1$ (squares), AD3, $\Delta\log(M/M_{\odot}) = 0.25 $ (circles), FF1, $\Delta\log(M/M_{\odot}) = 0.1$ (diamonds), and FF1, $\Delta\log(M/M_{\odot}) = 0.25$ (triangles). Average slopes computed over a minimum of 2000 samples. Open symbols indicate the slopes of the full sample size. }
\label{avgslope}
\end{figure}
\end{center}

\subsection{Effect of sample size and bin size on CMF slopes}

The main piece of data generally extracted from a CMF is the slope of the high mass tail. This information is then used to compare different regions to each other, and to the initial mass function (IMF) in an attempt to determine the true nature of star formation and the possible relation between the CMF and IMF. However, as discussed above, the sample size and bin size have a profound effect on the shape of the curve. This effect translates over to the derived slopes. To determine the extent of this effect, we generate multiple small sample CMFs (2000+) for each sample size and compute the average slope. Figure~\ref{avgslope} shows the results of this analysis for models FF1 and AD3 for both mass bin sizes. The filled symbols show the average slope for each of the models while the open symbols depict the slope of the full sample (10$^{6}$). Tests with larger numbers of samples for each sample size showed differences in the average slope of up to 0.01, which is encompassed in the size of the symbols.

As shown in Figure~\ref{avgslope}, the size of the bin clearly affects the average slope. The larger bin size yields slopes that are steeper than the Salpeter slope, while the smaller bin size shows an overall shallower average slope.  The size of the sample also effects the slope. Smaller samples generally result in steeper slopes than those derived using the full sample. 
Furthering this analysis we look at both the minimum and maximum slopes calculated for 
\begin{center}
\begin{figure}
\centering
\includegraphics[scale=0.15,angle=-90]{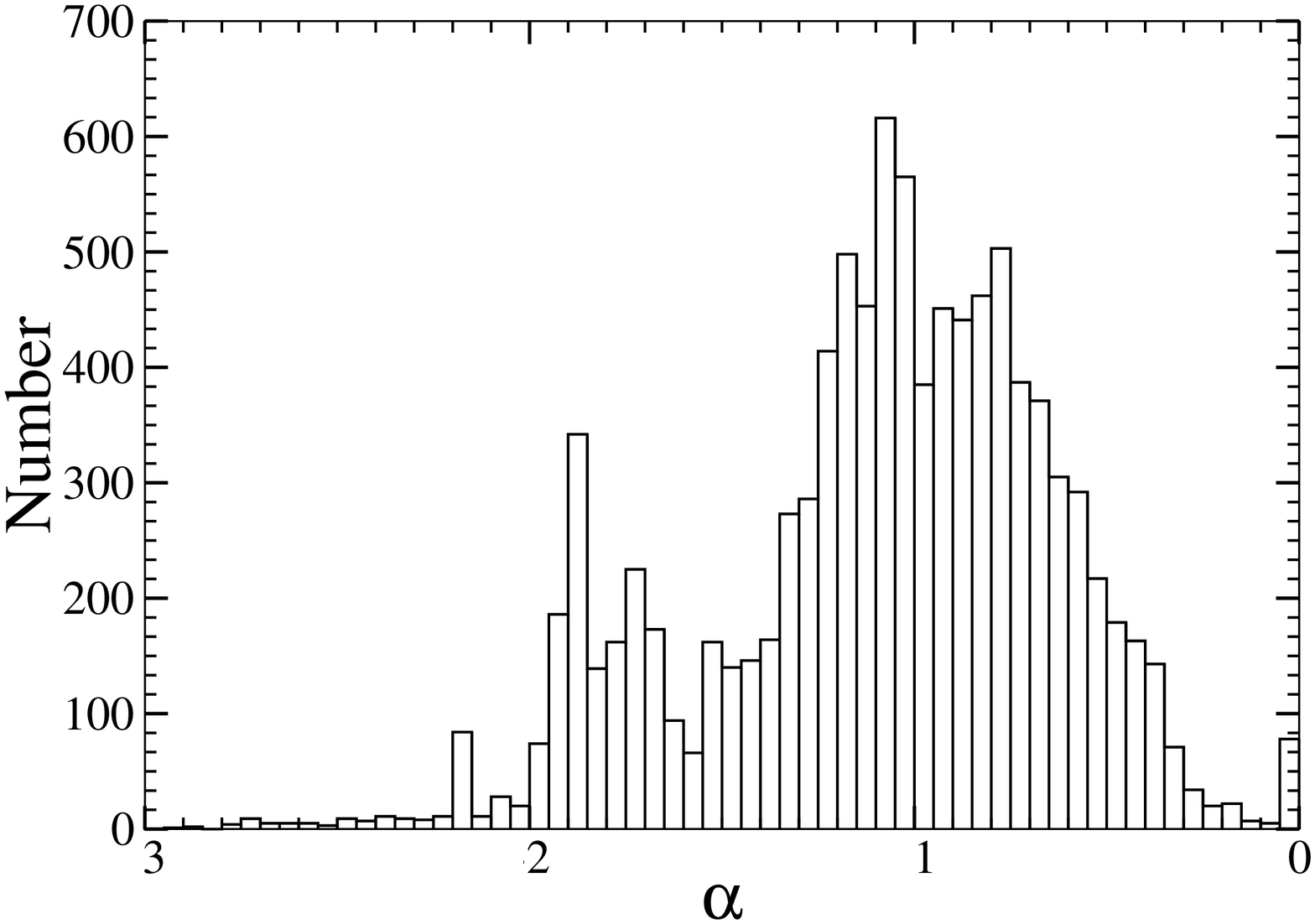}
\includegraphics[scale=0.15,angle=-90]{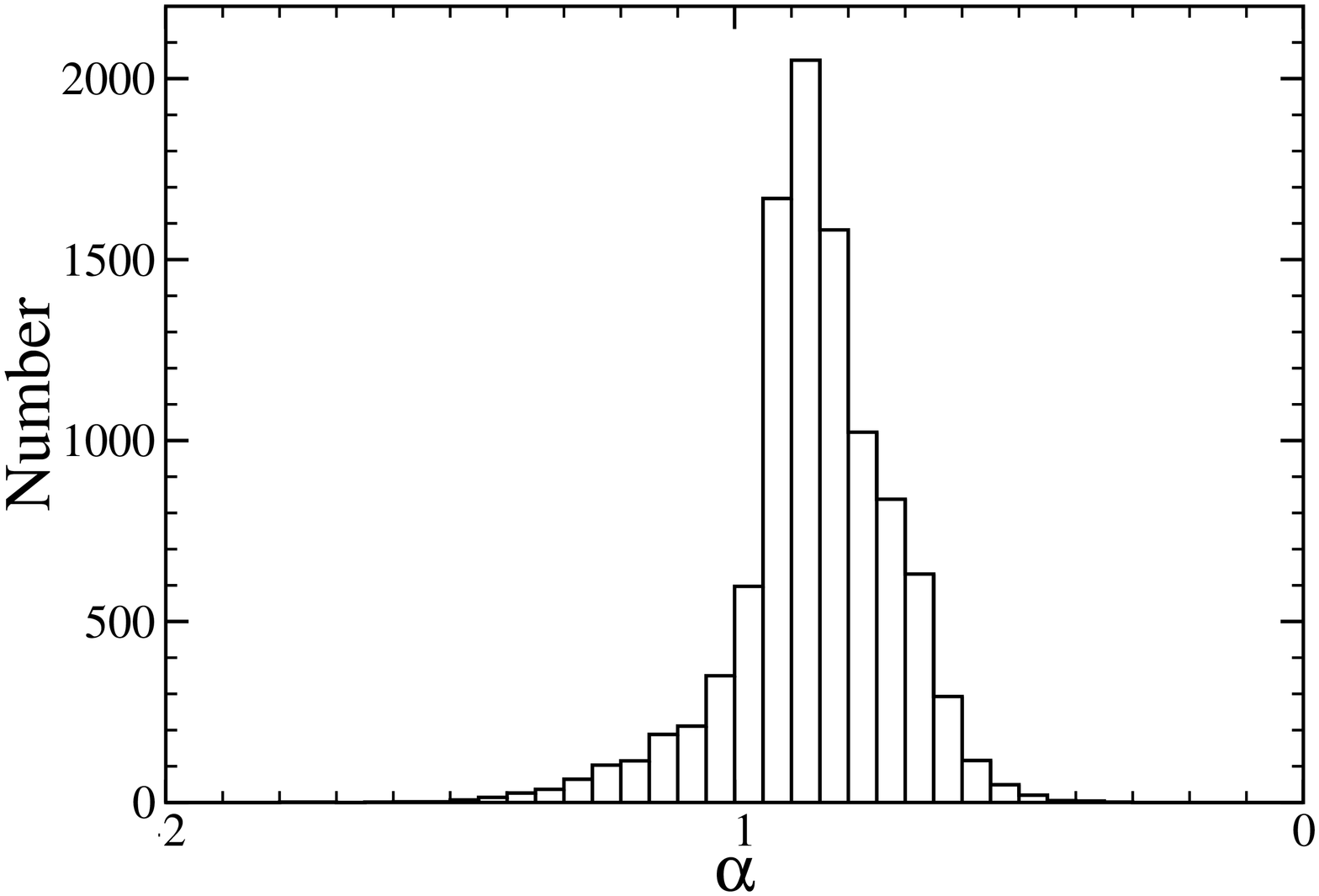}
\includegraphics[scale=0.15,angle=-90]{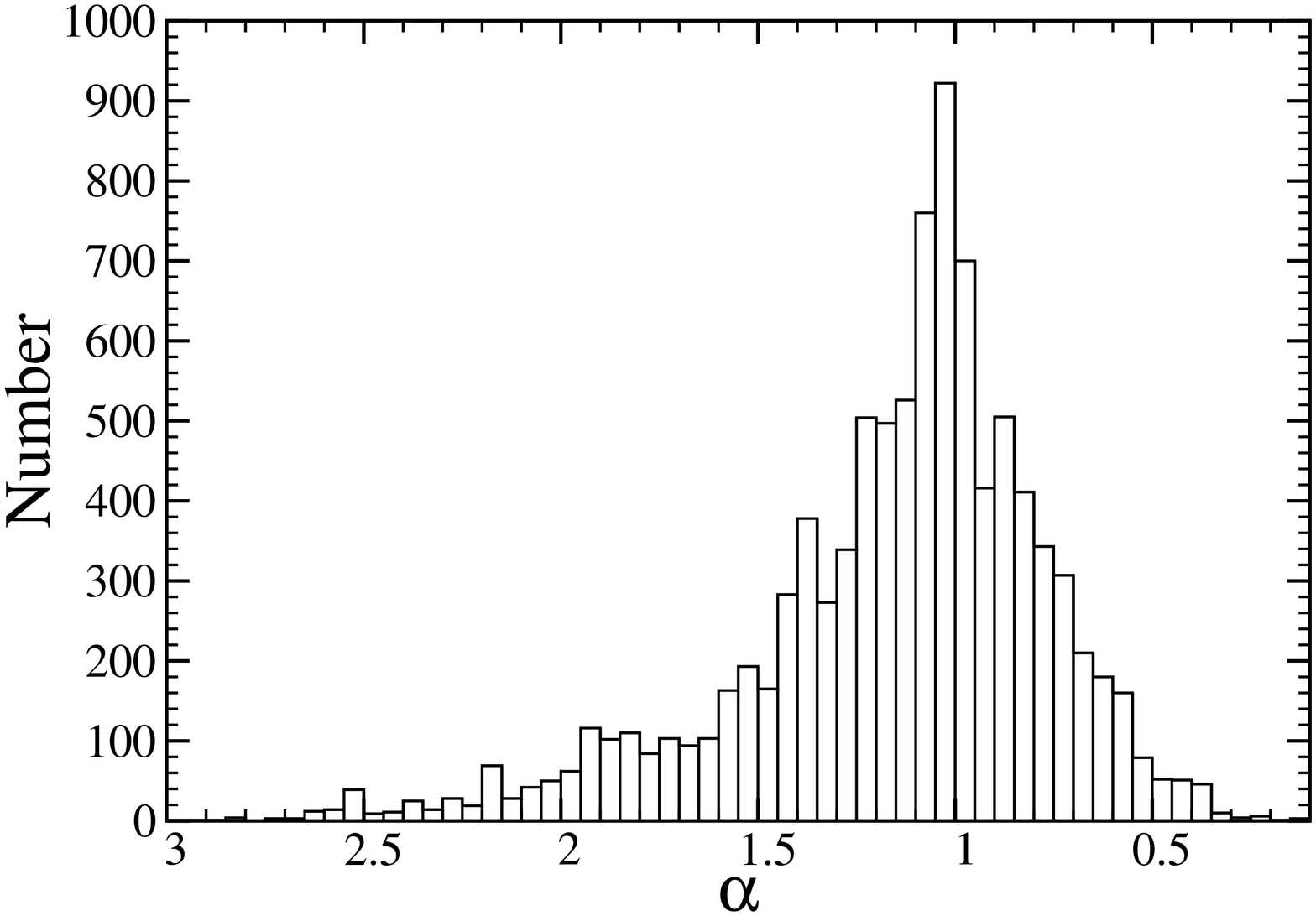}
\includegraphics[scale=0.15,angle=-90]{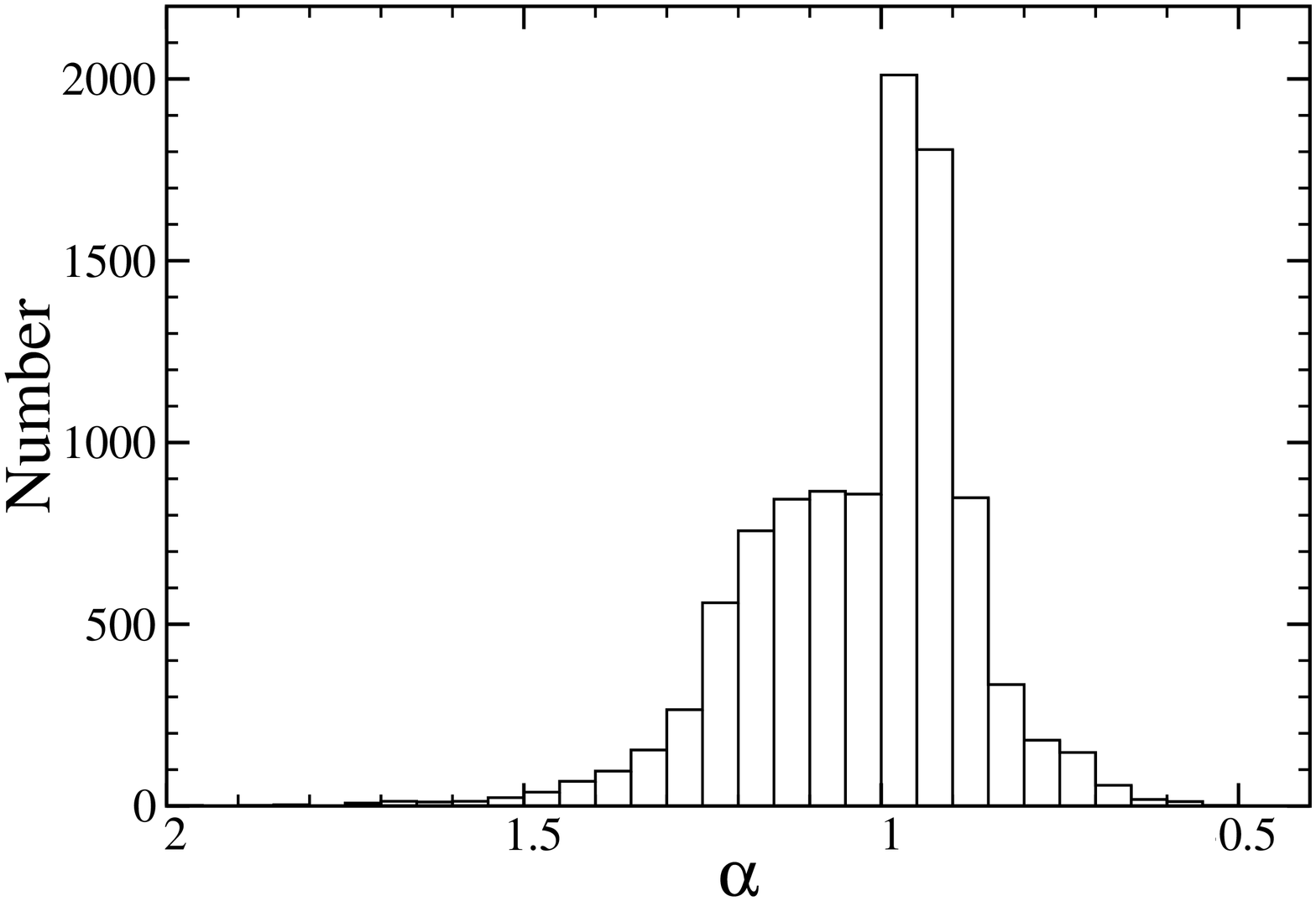}
\caption{Histogram of computed slopes for different models and sample sizes assuming $\Delta\log(M/M_{\odot}) = 0.1$ bins. Top row: AD3 with 100 cores per sample (left) and 500 cores per sample (right). Bottom row: Same as the top but for FF1.}
\label{slopehist}
\end{figure}
\end{center}
\begin{center}
\begin{figure}
\centering
\includegraphics[scale=0.15,angle=-90]{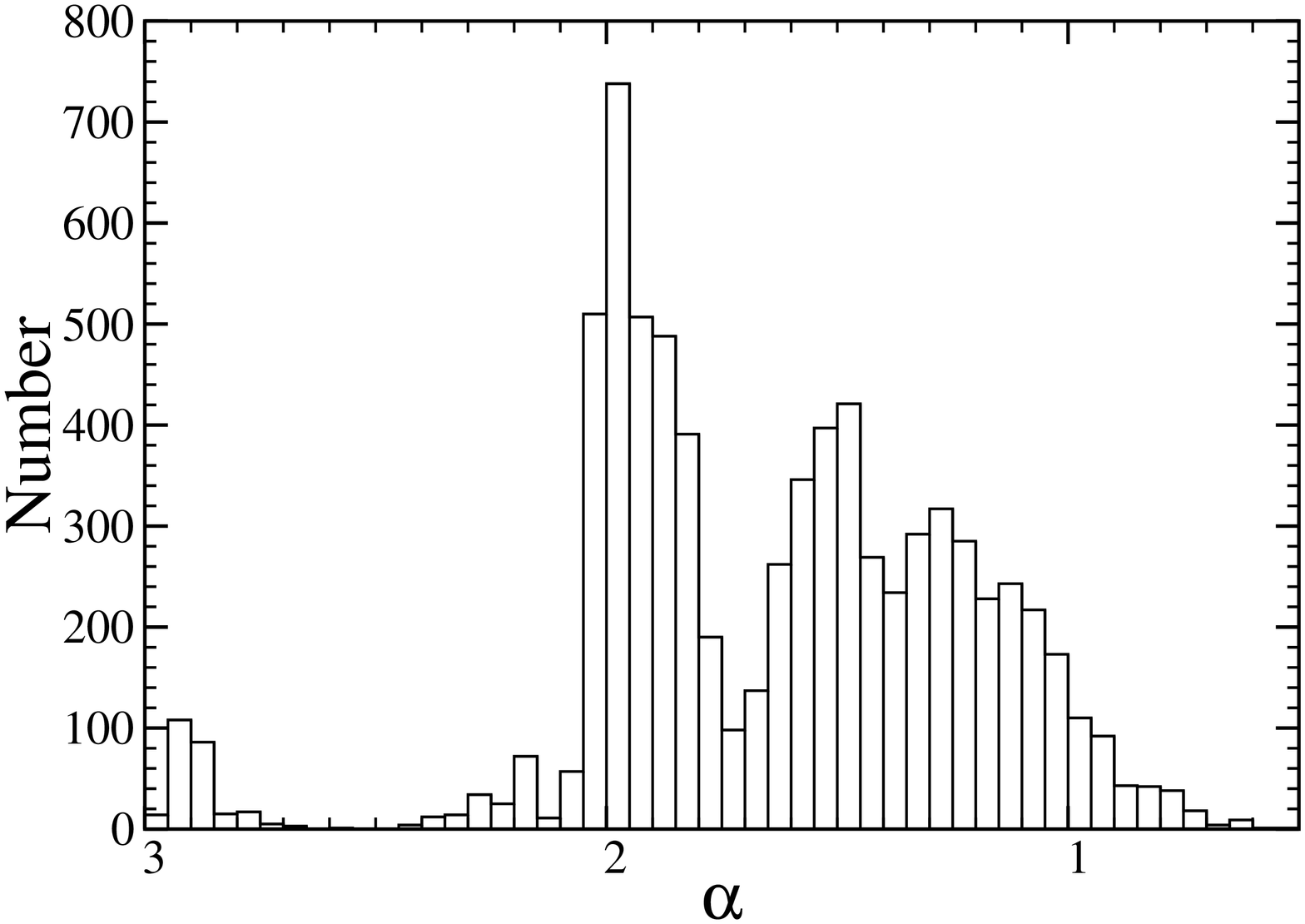}
\includegraphics[scale=0.15,angle=-90]{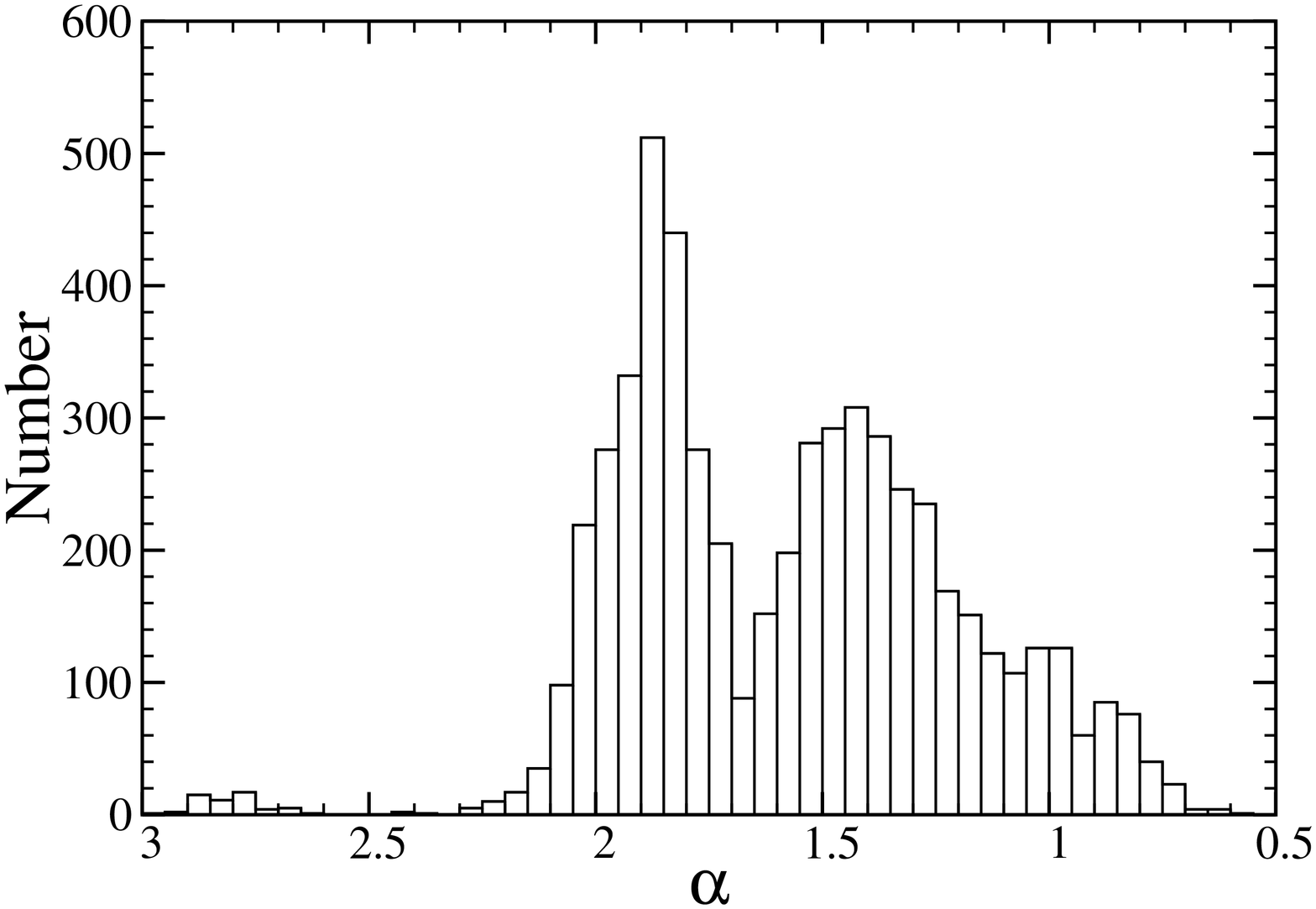}
\caption{Histogram of computed slopes assuming a CMF with 100 cores and $\Delta\log(M/M_{\odot}) = 0.25$ bins. Left: AD3, Right: FF1.}
\label{slopehist025}
\end{figure}
\end{center}
each  filled point in Figure~\ref{avgslope}, as well as the distribution of slopes. Figure~\ref{slopehist} shows the distribution of slopes for four of the points on Figure~\ref{avgslope} as indicated (Top row: AD3, Bottom Row, FF1. Left column: 100 cores, Right column: 500 cores) assuming a CMF constructed with $\Delta\log(M/M_{\odot}) = 0.1$ bins. All four cases show that the preferred slope value is close to the average slope value. The maximum and minimum computed slopes exhibit a very wide range for the small sample sizes (i.e., $\alpha = -0.1$ to $\alpha = 11$ for AD3, 0.1, 100 cores) while the larger sample sizes exhibit a smaller maximum-minimum range (i.e.,  $\alpha = 0.34$ to $\alpha = 1.79$ for AD3, 0.1, 500 cores). This decrease in the slope variance is evident when comparing the left column to those in the right column in Figure~\ref{slopehist}. From these plots, we conclude that although there can be a wide variance in possible slope values, the preferred slope value is in general smaller than the typical Salpeter value, $\alpha = 1.35$, and the range of slopes decreases as the number of samples increases. Figure~\ref{slopehist025} shows the distribution of slopes for two CMFs (left: AD3, right: FF1) assuming a 100 core sample size and $\Delta\log(M/M_{\odot}) = 0.25$ bins. Comparing to the left hand plots in Figure~\ref{slopehist}, we see that the larger bin size results in a bimodal distribution with the peaks occurring at $\alpha \approx 2.0$ and $\alpha \approx 1.5$ for AD3 and $\alpha \approx 1.9$ and $\alpha \approx 1.45$ for FF1. The result of this bimodal distribution is to shift the average slope values to smaller values than the dominant peak. This is particularly evident in Figure~\ref{avgslope} in the trend of slopes for the smallest sample sizes of the blue triangles (FF, 0.25). Further analysis of the effect of the original column density distribution on the variance and mean of the resulting slope histogram showed that a larger variance in the column density distribution shifts the mean in the slopes to smaller values ($< 1$) while a smaller variance results in a larger mean value, $\alpha \sim 1.35$.

\vspace{5mm}
\section{Discussion}

Our analysis shows that the shape of the CMF is highly dependent on the magnetic field strength and neutral-ion coupling within the cloud. Specifically, a flux-frozen magnetic field broadens the nonmagnetic lognormal distribution to have a significant power-law high mass tail, though it is much shallower than the Salpeter value. When ambipolar diffusion is taken into account, there is an intermediate mass tail and a high mass cutoff.  The extent of all these features are dependent on the range of mass-to-flux ratio values in the initial cloud.

\citet{KM2009} carried out a more focused study of the effect of magnetic fields and ambipolar diffusion in creating a broad CMF. Their model explored only the subcritical portion of the fragmentation scales seen in Figure~\ref{lammufig}. Furthermore, they assumed a uniform distribution of subcritical mass-to-flux ratios and effectively a fixed Jeans mass in order to generate their mass distribution.

The low-mass tail in their distribution originates in the assumption that the subcritical clouds ultimately form dense cores with masses that are scaled by $\mu_{0}$ for subcritical values of $\mu_{0}$. This is because numerical simulations of \citet{BM1995} show that only an inner region where the mass-to-flux ratio exceeds the critical value undergoes rapid collapse. We do not make that assumption in this study, since cores that form by ambipolar drift have an appearance that is similar to those that are forming by a more rapid gravitationally-dominated process \citep[see][]{Basu2009b}. Since the resultant CMF in our model is generated from an underlying lognormal function, it has an intrinsic 
peak even when binned in linear  mass bins. An advantage of the KM09 model is that they do not need to assume an underlying lognormal distribution to obtain a lognormal-like CMF, however that CMF is peaked only when binned in log mass.

Upon scaling our models down to observational sample sizes and ranges, we found that the distinction between the different models is lost within typical observational mass ranges and therefore no information regarding the magnetic field can be reliably gleaned from the shape of the observed CMFs. Further to this, analysis of the slopes for each of the sample sizes showed that the smaller samples sizes result in slopes that are $1.1~-~1.4$ times larger than the slope derived from the full sample, while the derived slopes for the larger binsize are $\sim1.3~-~2.0$ times larger than the corresponding smaller binsize slope measurements. Although we have taken care to scale our analysis down to those typically used in observations, the question still remains as to how well our results and conclusions correspond to actual observations. A recent study of the CMF for five separate star forming regions (Ophiuchus, Taurus, Perseus, Serpens and Orion) performed by \citet{Sadavoy2010} provides the perfect platform for comparison. Looking at the core mass distributions for these regions, as expected, it is hard to definitively discern any characteristic features that are indicative of a particular magnetic field model. With limited data, it is plausible that the CMFs for Ophiuchus, Taurus and Perseus could exhibit the indicative shoulder of the ambipolar diffusion models, while the full Orion CMF could show evidence of a flux frozen field. Looking at the slopes of the CMFs for these regions, \citet{Sadavoy2010} showed  each region gave slope values that are close to the $\alpha = 1.35$ Salpeter slope, within their adopted errors. Comparing their slope values to those in Figure~\ref{avgslope}, most of them would fall somewhere in the lower half of the graph in amongst the diamonds and squares while the Orion with OMC slope would fall in amongst the triangles and circles. However looking at the binsize of the observations, all of the slopes should be within the triangle/circle regime of the graph.  Comparing these values to the corresponding slope histograms (see Figure~\ref{slopehist025}) we see that these values all fall within the regime of possible slopes. On the surface, this seems to be a huge discrepancy between our results and observations, however each of these five observational slopes represents a single slope within our 2000+ values used to derive an average slope. However, looking at the range of slopes derived from our analysis, these observed slopes fall within this range. As shown in Figure~\ref{slopehist}, the only way to produce a narrower range of slope values is to increase the sample size, which is not always possible observationally since the number of objects detected depends entirely upon the number of objects actually present and the sensitivity of the instrument.

Based on our analysis and the above comparison to the work by \citet{Sadavoy2010}, we argue that the observed CMFs are extremely statistically limited, both in the size of the sample and the number of samples over which the slope of the CMF is averaged. Through our analysis, we have shown that with larger number statistics, not only is the measured slope of the CMF much different than the typical Salpeter value $\alpha = 1.35$, but highly dependent on the size of the mass bin. In addition, the range of individual slope values within the set size decreases as the number of cores in the sample size increases.  This is analogous to the results found by \citet{Elmegreen1999}, where although it was determined that the most probable value for the IMF slope is the Salpeter value,  $\alpha = 1.35$,  it is a highly reduced average of all possible outcomes. Subsequently, we argue that based on our analysis and the results of \citet{Elmegreen1999}, there seems to be no clear cut correlation between the slope of the CMF and the IMF and that the shape and slope of the CMF are entirely controlled by the conditions within the cloud itself. Since it is unfeasible to claim that all clouds exhibit identical conditions, it is therefore unrealistic to expect a universal shape and slope value for all star forming regions. 

\section{Summary}
We have studied the effect of magnetic fields on the formation and properties of the core mass function using a combination of the results from linear analysis and Monte Carlo methods. In addition, we have studied the effects of low number statistics on the slope of the high mass tail. Here we summarize the main results of our analysis.

\begin{itemize}
\item The synthetic CMFs show that the presence of a magnetic field has several effects on the shape of the CMF. In general, a magnetic field acts to broaden the core mass function compared to the nonmagnetic CMF. In addition, the magnetic CMFs exhibit a high mass tail. The form of this tail depends on whether the field is flux frozen or allows for neutral-ion drift across the field lines. In the former case, the tail exhibits a continuous power law while in the latter case, the high mass tail truncates to form a ``shoulder''.

\item The nonmagnetic model shows that the high mass cores are formed from low density gas and vice versa. Analysis of the contributions of low and high density gas to the low and high mass regions of the CMF shows that the addition of magnetic fields results in additional contributions of high mass cores formed from high density gas.

\item Scaling of the synthetic CMFs down to typical observational sample sizes and bin sizes show that the ability to distinguish between the different models is no longer possible for the smallest sample sizes (100 cores) and typical bin sizes ($\Delta\log(M/M_{\odot}) = 0.25$). This shows that the current observations of core mass functions are statistically limited.

\item Statistical analysis of the derived slope from a large sample of synthetic CMFs show that the slope of the high mass tail is systematically steeper for smaller core sample sizes than for larger sample sizes. In addition, the average slope is also systematically steeper for larger bin sizes. 

\item Analysis of the minimum, maximum and distribution of calculated slopes shows that the most probable slope does not necessarily correspond to the canonical Salpeter value. In addition, the most probable slope value becomes shallower as the sample size increases. 
\end{itemize}

\section*{Acknowledgments}
NDB was supported by a scholarship from the Natural Science and Engineering Research Council (NSERC) of Canada. SB was supported by a Discovery Grant from NSERC.


\end{document}